\documentclass[aps,prb,groupedaddress,twocolumn, amsmath,amssymb,showpacs,
eqsecnum,floatfix]{revtex4}

\usepackage{graphicx}
\begin{document}


\title{Phase diagram of vortex matter in layered superconductors with random
point pinning}


\author{Chandan Dasgupta}
\email{cdgupta@physics.iisc.ernet.in}
\altaffiliation{Also at Condensed Matter Theory Unit, Jawaharlal Nehru Centre
for Advanced Scientific Research, Bangalore 560064, India}
\affiliation{Centre for Condensed Matter Theory, Department of Physics, 
Indian Institute  of Science, Bangalore 560012, India}
\author{Oriol T. Valls}
\email{otvalls@umn.edu}
\altaffiliation{Also at Minnesota Supercomputer Institute, University of Minnesota,
Minneapolis, Minnesota 55455}
\affiliation{School of Physics and Astronomy,
University of Minnesota, Minneapolis, Minnesota 55455}

\date{\today}

\begin{abstract}
We study the phase diagram of the superconducting vortex system in 
layered high-temperature superconductors
in the presence of a magnetic field 
perpendicular to the layers and of random atomic scale point pinning
centers. We consider the highly anisotropic limit where the pancake vortices
on different layer are coupled only by their electromagnetic interaction.
The free energy of the vortex system is then represented as a 
Ramakrishnan-Yussouff free energy
functional of the time averaged vortex density. We numerically minimize
this functional and examine the properties of 
the resulting phases. We find that, in the temperature ($T$) -- pinning strength
($s$) plane at constant magnetic induction, the equilibrium phase at low 
$T$ and $s$  is a Bragg glass. As one increases $s$ or $T$ a first order
phase transition occurs to another phase that we characterize as
a pinned vortex  liquid. The weakly pinned 
vortex liquid obtained for high $T$
and small $s$ smoothly crosses over to the strongly pinned vortex liquid
as $T$ is decreased or $s$  increased -- we do not find
evidence for the existence, in thermodynamic
equilibrium, of a distinct vortex glass phase in the range of pinning 
parameters considered here. 
We present results for the density correlation
functions, the density and defect distributions, and the local 
field distribution
accessible via $\mu$SR experiments. These results are compared with those of
existing theoretical, numerical and experimental studies.
\end{abstract}


\maketitle

\section{Introduction}
\label{intro}

Equilibrium and transport properties of the mixed phase of 
highly anisotropic, layered, high-temperature
superconductors (HTSCs) in the presence of various kinds of 
pinning have been studied
extensively for over one decade.\cite{review} Random pinning destroys the
long-range translational order of the Abrikosov lattice and leads to a variety
of low-temperature glassy phases.  
In systems with random point pinning, the existence of a topologically
ordered Bragg glass (BrG) phase with quasi-long-range translational
order at low temperatures, low fields and weak pinning
is now well-established both theoretically~\cite{natt,giamarchi}
and experimentally\cite{klein}. The BrG is expected to melt into a vortex
liquid (VL) via a first-order transition as the temperature is increased.
The possibility of occurrence of an amorphous
vortex glass (VG) phase (with nonlinear voltage-current
characteristics and vanishing resistance in the zero-current limit) 
at low temperatures in
systems with strong pinning (or at high magnetic fields where the effects
of pinning disorder are enhanced) was suggested 
several years ago~\cite{mpa,ffh}.
However, in spite of extensive
investigation over almost twenty years,
whether a true VG phase (that is, an amorphous glassy phase 
thermodynamically distinct
from the high-temperature VL) exists in systems with uncorrelated
point pinning remains controversial.

Evidence for the theoretically expected scaling behavior of the 
current-voltage characteristics near
a continuous VL--VG transition was obtained in several early 
experiments.\cite{vglexp} The validity of the conclusions drawn from
these experiments has been questioned in latter studies,\cite{lobb} 
although a recent experimental study\cite{zeldov} 
has presented intriguing thermodynamic evidence for
the existence of a VG phase.
Numerical studies\cite{bokil} of a ``gauge-glass'' model  
believed to be in the
same universality class as vortex systems with random point
pinning suggest that a finite-temperature VG phase does not 
exist in three dimensions if electromagnetic screening of the interaction
between vortices is taken into account. 
Other numerical studies\cite{olsson,lidmar,nh} of models in which the 
interaction between vortices is not screened indicate
the presence of a thermodynamically stable VG phase at low temperatures 
if the pinning is strong.
One of these\cite{nh} also presents evidence for the existence of
a ``vortex slush'' phase with short-range translational order (i.e. a
translational correlation length significantly longer than that in the
VL phase) but no
long-range phase coherence. However, the existence of this phase has been
questioned in later work.\cite{teitel} 
A variety of interesting ``glassy'' behavior has been experimentally observed
\cite{banerjee} near the first-order
melting transition of the BrG phase of superconductors  (both
conventional and
HTSCs) with point pinning. It is possible that\cite{banerjee,gautam}
these observations can be understood by assuming that the
melting of the BrG phase occurs in two steps: the BrG first transforms into
a ``multidomain'' glassy phase that then melts into the usual VL at a
slightly higher temperature. 

It is clear from this brief summary of the many existing 
theoretical and experimental results that several important issues about the
structure and phase behavior of vortex matter in the presence of random point
pinning are still unresolved.

In this paper we present the results of a numerical study of the structural
and thermodynamic properties of the system of vortices in a highly anisotropic
layered superconductor in the presence of random point pinning. The magnetic
field is perpendicular to the superconducting layers and 
the vortex lines can then be considered as
stacks of pancake vortices residing on
the layers. These pancake vortices constitute
a system of point-like objects that interact among themselves and also with
an ``external'' potential arising from a small
concentration of randomly placed pinning centers (atomic scale
defects). The 
positions of these defects on different layers are assumed to be completely
uncorrelated. In this
highly anisotropic case, the 
Josephson coupling between layers is vanishingly small, and 
pancake vortices on different layers are coupled
only through the electromagnetic interaction. 
Under these circumstances, one can use (as 
in our recent studies~\cite{prbv,prbd} of vortex systems in the presence of
columnar pinning)   a 
spatially discretized
version of the Ramakrishnan-Yussouff free energy functional~\cite{ry} for
a system of pancake vortices. In this formalism, the free energy of the
system is expressed as a functional of the time-averaged local density of 
vortices. The different local minima of this
functional can then be found by starting the numerical
minimization process with appropriate
initial conditions.  These local minima, whose density
configurations are obtained in our calculations, 
can then be identified with different 
phases of the system in our mean field description: the nature of the phase
corresponding to a particular local minimum is determined
from a detailed analysis of the density distribution and the
correlation functions at that minimum. A
first-order transition between two phases, represented by two different
locally stable minima of the free energy functional, is signaled by a
crossing of the free energies of these minima as some control parameter
(e.g. the temperature) is varied.

Using this method, we have analyzed the phase diagram of the system as a 
function of temperature and pinning strength for a fixed value of the 
magnetic induction, which determines the areal density of pancake vortices.
For low temperatures and relatively small values of the pinning strength,
we find a nearly crystalline minimum that exhibits all the properties expected
for a BrG phase (the ``BrG minimum''). This 
BrG phase is the thermodynamically stable one (has the lowest
free energy) if the
temperature is sufficiently small and the pinning sufficiently weak. It
becomes unstable very soon after either of these parameters is increased
beyond the region where this is the equilibrium phase. We
find another local minimum at which the density distribution in each layer
is strongly 
disordered (amorphous) and the densities in different layers are weakly 
correlated. This minimum is at least locally stable for  the range of temperature
and pinning strength we have considered. If the temperature is high and
the pinning strength low, then the density distribution at this minimum is
weakly inhomogeneous, showing the characteristics 
expected for a weakly pinned VL. As the temperature is reduced, or the pinning
strength is increased, this minimum evolves continuously into a state with
strong inhomogeneity in the distribution of the time-averaged local density.
This state represents a system of pancake
vortices strongly localized at random positions because
of the random pinning potential. This state exhibits  characteristics of
a glass, but we prefer to call it a strongly pinned VL because it is 
continuously connected to the weakly pinned VL found for high temperatures and 
weak pinning --  we do not find any indication of a phase transition between
the liquid-like and glass-like states. 

As the temperature is increased at a fixed, relatively small value of
the pinning strength, or the pinning 
strength is increased at a sufficiently low temperature, 
the free energy of the BrG minimum crosses that of the VL minimum, signaling
a first-order phase transition from the BrG to the pinned
VL phase. For very weak
pinning, the temperature at which this transition occurs is nearly independent
of the pinning strength: it is very close to the melting temperature of the 
unpinned vortex 
lattice. As the pinning strength is increased, the
transition temperature decreases and eventually, the transition line in the
temperature -- pinning strength plane exhibits a trend 
towards becoming parallel to the
temperature axis. Thus, the BrG phase is 
thermodynamically stable only if the pinning strength is lower than a critical
value. The shape of the boundary of the BrG phase is 
similar to that found in existing numerical simulation 
studies.\cite{nh,teitel2}
As mentioned above, the local minimum corresponding to the VL
phase remains locally stable as the transition line to the BrG phase is crossed
either by reducing the temperature or by decreasing the pinning strength. On 
the other hand, the BrG minimum becomes unstable (and the minimization 
converges to a minimum of the VL type) as the temperature or the pinning 
strength is increased slightly beyond the point at which the transition to the
VL phase occurs. This behavior is consistent with experimental 
results~\cite{andrei} that indicate that 
the superheated BrG phase in ${\rm NbSe_2}$ exhibits a spinodal instability, whereas 
there is no limit of metastability of the supercooled disordered state.   
We have also carried out a careful search for the 
presence of local minima of the free energy that
may correspond to the ``vortex slush'' phase~\cite{nh} or the polycrystalline
``domain glass'' phase~\cite{gautam}  
predicted in some earlier studies. As we will explain
below, we did not find any
 equilibrium
free energy minimum that could be interpreted as either one of these phases
although we have found some indications that they may occur under
nonequilibrium conditions.

We have used our results for the vortex density distribution
at the free energy minima to calculate
the distribution of the local (microscopic) magnetic induction in the sample.
This is important because the distribution of the local magnetic
induction is measured in muon spin rotation ($\mu$SR) experiments, 
and phase transitions in the vortex system give rise to characteristic changes
in the width and shape of this distribution. Thus, $\mu$SR experiments have been
widely used~\cite{musrrev} to study phase transitions in the mixed state of
HTSCs and conventional type-II superconductors. 
Our results show a relatively broad and asymmetric distribution of the
microscopic magnetic induction in the BrG phase, and a narrow and much more
symmetric distribution in the VL phase. These features 
are in good
agreement with available $\mu$SR data.  

The rest of this paper is organized as follows. In section~\ref{methods}, we
define the model we consider and describe the numerical method
used in our study. The results of our calculations are described in detail
and compared with existing experimental and numerical results in 
section~\ref{results}. Section~\ref{summary} contains a summary of our main 
results and a few concluding 
remarks.

\section{Model and Methods}
\label{methods}

We consider a  layered superconductor
in a magnetic field 
perpendicular to the layers. We assume that the material is strongly
anisotropic
so that the  Josephson coupling between layers is vanishingly small.
The pancake vortices on different layers are then 
coupled only through their electromagnetic interaction. In this 
limit, vortices
in the same layer repel each other with a potential logarithmic
in their separation, while there is
a much weaker attractive interaction between vortices on different layers
which falls off exponentially with layer separation and varies
logarithmically with the separation in the layer plane.
This model for the vortex system was used
in our earlier studies~\cite{prbv,prbd} of the effects of columnar 
pinning. 
As discussed in detail in Ref.~\onlinecite{prbv}, many theoretical
and experimental studies indicate that this model is appropriate for describing
the properties of vortex matter in highly anisotropic materials such as BSCCO 
(${\rm Bi_2Sr_2CaCu_2O_{8+x}}$) in a large region of the field--temperature
plane, although this conclusion has been questioned
in a recent theoretical analysis~\cite{roz}.

The free energy functional of such a system of pancake vortices
lying on the superconducting layers and interacting with a time-independent
pinning potential can be written in the form:
\begin{equation}
F[\rho]=F_{RY}[\rho]+F_p[\rho]
\label{fe}
\end{equation}
The first term
in the right side of Eq.~(\ref{fe}) is the free energy of the vortex system
in the absence
of pinning, while the second includes the
pinning effects. The first term is of the Ramakrishnan-Yussouff 
form~\cite{ry}:
\begin{widetext}
\begin{equation}
\beta F_{RY}[\rho] = \sum_{n}\int{d {\bf r}\{\rho_n({\bf r})
\ln (\rho_n({\bf r})/\rho_0)-\delta\rho_n({\bf r})\} } 
 -(1/2)\sum_m \sum_n \int{d {\bf r} \int {d{\bf r}^\prime
C_{mn}({|\bf r}-{\bf r^\prime|}) \delta \rho_m ({\bf r}) \delta
\rho_n({\bf r}^\prime)}} ,
\label{ryfe}
\end{equation}
\end{widetext}
where $\beta$ is
the inverse temperature. We have defined
$\delta \rho_n ({\bf r})\equiv \rho_n({\bf r})-\rho_0$ as the
deviation of  ${\rho_n(\bf r})$, the time-averaged areal density of
pancake vortices at point $\bf r$ on layer $n$, from $\rho_0$,
the density of the uniform liquid ($\rho_0 = B/\Phi_0$ where $B$ is the
magnetic induction and $\Phi_0$ is the flux quantum) and
we have taken our zero of the free energy at its uniform liquid value.
The integrations are two-dimensional and the sums are over all layers.
The function $C_{mn}(r)$, which depends on the layer separation $|m-n|$
and the separation $r$ in the layer plane, is the direct pair correlation
function of the layered vortex liquid~\cite{hansen}
at density $\rho_0$. This static
correlation function
contains all the required information about the interactions in the system.
We have used here the $C_{mn}(r)$ obtained from a 
calculation~\cite{menon1} via the 
hypernetted chain approximation~\cite{hansen} for parameter values appropriate
for BSCCO, 
as in Ref.~\onlinecite{prbd}: $d=15 \AA$ for the interlayer distance,
and a two-fluid form for the temperature dependence 
of the penetration depth 
$\lambda(T)$ with $\lambda(0)=1500 \AA$. 

The second term in Eq.(\ref{fe}) represents the effects of pinning and
can be written as
\begin{equation}
F_p[\rho]= \sum_n \int{d {\bf r} V^p_n({\bf r}) [\rho_n({\bf r})-\rho_0] },
\end{equation}
where $V^p_n({\bf r})$ is the pinning potential
at point ${\bf r}$ on layer $n$.
The pinning potential is assumed to be produced by
random atomic scale point defects. The
positions of the defects on different layers are assumed to be completely
uncorrelated. Each defect is assumed to produce a pinning potential of depth
$v_0$ and having a short range which we will take
(see below) as of the order of the spacing $h$ 
of the computational mesh in our numerical calculation.
A randomly chosen small fraction, $c$, of the unit cells of the
{\it underlying crystal lattice}, taken to be
a square lattice with spacing $d_0 = 4\AA$, are assumed to be occupied by
pinning defects. 

If the
time averaged vortex density is the same on all layers (as would be the case
for columnar pins perpendicular to the layers), the free energy could be
written in an effectively two-dimensional form~\cite{prbv,prbd} with the
quantity $\tilde{C}(r) \equiv
\sum_n C_{mn}(r)$ playing the role of the two-dimensional direct correlation
function. But in the present case, with a pinning potential that
varies from one layer to another, the full three dimensional problem has to
be considered.
 The problem is therefore computationally much more
demanding.
To numerically minimize the free energy of the system, we discretize the
density variable by introducing a layered triangular computational lattice 
containing $N^2$ sites on each layer. The total
number of computational mesh points is thus $N^2 N_L$ where $N_L$ is the
number of layers considered. We use periodic boundary conditions 
in all three directions.
On the sites of this lattice we define discretized density variables
$\rho_{n,i} \equiv \rho_n({\bf r}_i) A_0$, where $\rho_n({\bf r}_i)$ is the
density at mesh point $i$ on layer $n$ and $A_0$ the area of the unit
cell of the in-plane computational lattice. A numerical procedure developed
earlier\cite{prbv,cdo} is used to find local minima of the free
energy written as a function of the variables
$\{\rho_{n,i}\}$.

We use the quantity $a_0$ defined by $\pi\rho_0a_0^2 \equiv 1$ as our
unit of length.
The spacing of the triangular computational lattice is taken to be
$h=a/16$ where $a=1.998 a_0$ is
the equilibrium lattice constant~\cite{prbv}
of the Abrikosov vortex lattice in the absence of pinning
in the temperature range considered (see below)
which is chiefly determined by the melting temperature of the
unpinned vortex lattice, namely $T_m^0 \simeq $ 18.4K~\cite{prbv}
at the field considered in this work ($B$ = 0.2T). 
The total number of defects on each layer of our sample is
given by $N_d = \sqrt{3}(Nh)^2 c /(2 d_0^2)$. In our discretized system, the
values of the pinning potential $V_{n,i}$ associated with mesh point $i$ on the
$n$th layer are determined in the following way: for each layer, we generate
$N_d$ points distributed randomly and uniformly over the area of the layer.
The potential $V_{n,i}$ is then taken to be $v_0 (m_{n,i}-N_d/N^2)$
where $m_{n,i}$
is the number of such point in the $i$th computational cell on layer $n$,
and $N_d/N^2$ is the average value of $m_{n,i}$.
We take $c$ to be 0.01 (1\%) and measure $v_0$ in units of $k_B$, the
Boltzmann constant, so that the parameter $s \equiv v_0/k_B$ (with values
given in Kelvins) measures the strength of the pinning potential. With
this parametrization, 
we have $\langle V_{n,i} \rangle = 0$, and $\langle
\beta V_{n,i} \beta V_{n^\prime,j} \rangle = 2.68 (s/T)^2 \delta_{nn^\prime}
\delta_{ij}$ where $T$ is the
temperature and the angular brackets denote the average over the random
configuration of defects.

\begin{figure}
\includegraphics [scale=0.4] {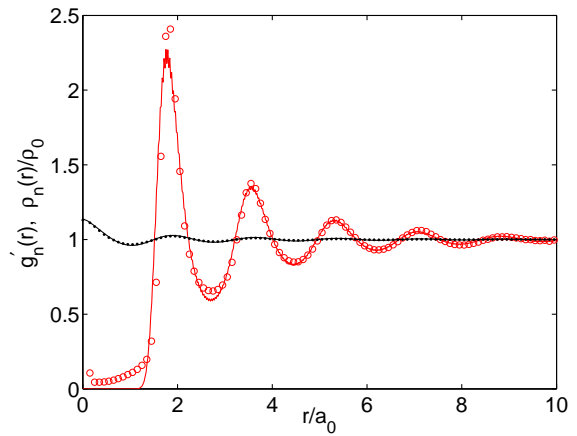}
\caption{\label{fig0} (Color online). Plots of the local density $\rho_n(r)$,
normalized by the average density $\rho_0$, for $n=0$ ((red) circles) and
$n=1$ ((black) dots), as functions of $r/a_0$. Here, $n$ is the layer
separation and $r$ is the in-plane distance from a single strong pinning center 
that traps a pancake vortex. The results for the pair distribution function
${g^\prime}_n(r)$ are also shown for comparison 
((red) dashed line: $n=0$; (black)
solid line: $n=1$). These results are for the vortex liquid state at temperature
$T$ = 20K and field $B$ = 0.2T.} 
\end{figure}

To test the numerical method described above, we carried out a minimization
of the free energy for a system in which a single strong pinning center was
placed on one of the layers. The depth and the range of the
potential produced by the pinning center were chosen (as in 
Ref.~\onlinecite{prbv}, see Fig.~3 in that work) so as to ensure that it traps a single 
vortex at the temperature $T$ = 20K (the unpinned vortex system is in the
liquid state at this temperature~\cite{prbv}). According to a
well-known result in liquid state theory~\cite{hansen}, the time-averaged 
local density $\rho_n(r)$ outside the range of the potential produced
by the pinning center (here, $n$ denotes the separation from the layer in
which the pinning center is located, and $r$ is the in-plane distance from the
position of the pinning center) should be equal 
to $\rho_0 {g^\prime}_n(r)$ where 
${g^\prime}_n(r)$ is the pair distribution function 
of the unpinned vortex liquid,
which can be obtained from the direct pair correlation function $C_n(r)$
used as input in our calculations. The results for $\rho_n(r)/\rho_0$, 
obtained from a numerical minimization of the discretized free energy
(with $N=128$ and $N_L=32$) are shown in Fig.\ref{fig0}
for $n$ = 0 and $n$ = 1 (for $n=0$, the results for $\rho_n(r)/\rho_0$ are
shown for values of $r$ outside the range of the potential produced by 
the pinning center at $r=0$). 
The plots also show the values of ${g^\prime}_n(r)$ obtained
from the $C_n(r)$ used as input. The agreement between the two sets of results
is quite good (the small differences may be attributed to the fact that the
peak of the local density at the pinning center at $r=0$ is not a 
$\delta$-function). These results confirm that the numerical
method used in our calculations correctly reproduces the strong intra-layer 
correlations, as well as the much weaker inter-layer correlations of 
local density of the system of pancake vortices.  

\section{Results}
\label{results}

We now
present our results. The values of the temperature
$T$ and pin strength parameter $s$ are indicated in
each  case. The field is always $B=0.2 T$. 
The number of layers is $N_L=128$ and the transverse computational lattice
size $N=256$. With the value of the computational lattice constant
$h$ being $h=a/16$ where $a=1.988 a_0$ is\cite{prbv} the equilibrium lattice
constant of the vortex lattice, the number of pancake vortices in each layer 
is $N_v=256$. The absence of finite size effects has been verified by
checking that these larger-size results are nearly
identical to those obtained for 
smaller systems.
Results have been obtained for five random 
pin configurations and  averaged over this number 
(except as indicated) whenever appropriate.
This is a  sufficient number as we find that the differences 
between results for
different pin configurations of the same strength and concentration
are negligible for the quantities of interest.

We use several kinds of initial conditions to start the minimization
process. In the first kind, we start with uniform ``liquid-like''
conditions,
with the variables $\rho_{n,i}$ all equal to their average value. The second
kind is ``crystal-like'':  the initial $\rho_{n,i}$ variables
are  obtained, for a given random pin configuration,
by minimizing the pinning energy of the equilibrium crystalline configuration
in the absence of pinning\cite{prbv} for the 
value of $B$  used here, with respect to the symmetry
operations of the computational 
lattice. Thirdly, once a local minimum configuration
has been obtained from the minimization procedure for certain values
of $s$ and $T$, that configuration is used as the initial condition
for a nearby value of either one of these variables. 
The first kind of initial conditions leads, after minimization, to
free energy minima which are, as we shall see below, disordered in
a liquid-like or glass-like  way. The second kind yields BrG
states, provided however that the values  of $T$ or $s$ are not too  high:
if they are, the crystal-like BrG state is not stable and those 
initial conditions
lead eventually to a disordered state. The third kind leads to a state of the
same general kind as the initial one, except when an instability boundary
is crossed. 

\subsection{Structure of free energy minima}

>From the set of values for the $\rho_{n,i}$ at each local minimum, the nature of
that minimum can be ascertained by evaluating the appropriate correlation
functions and also by direct visualization of either the density
variables or of the ``vortex lattice'' formed by the locations of the  
peaks of the local density. 

\begin{figure}
\includegraphics [scale=0.62] {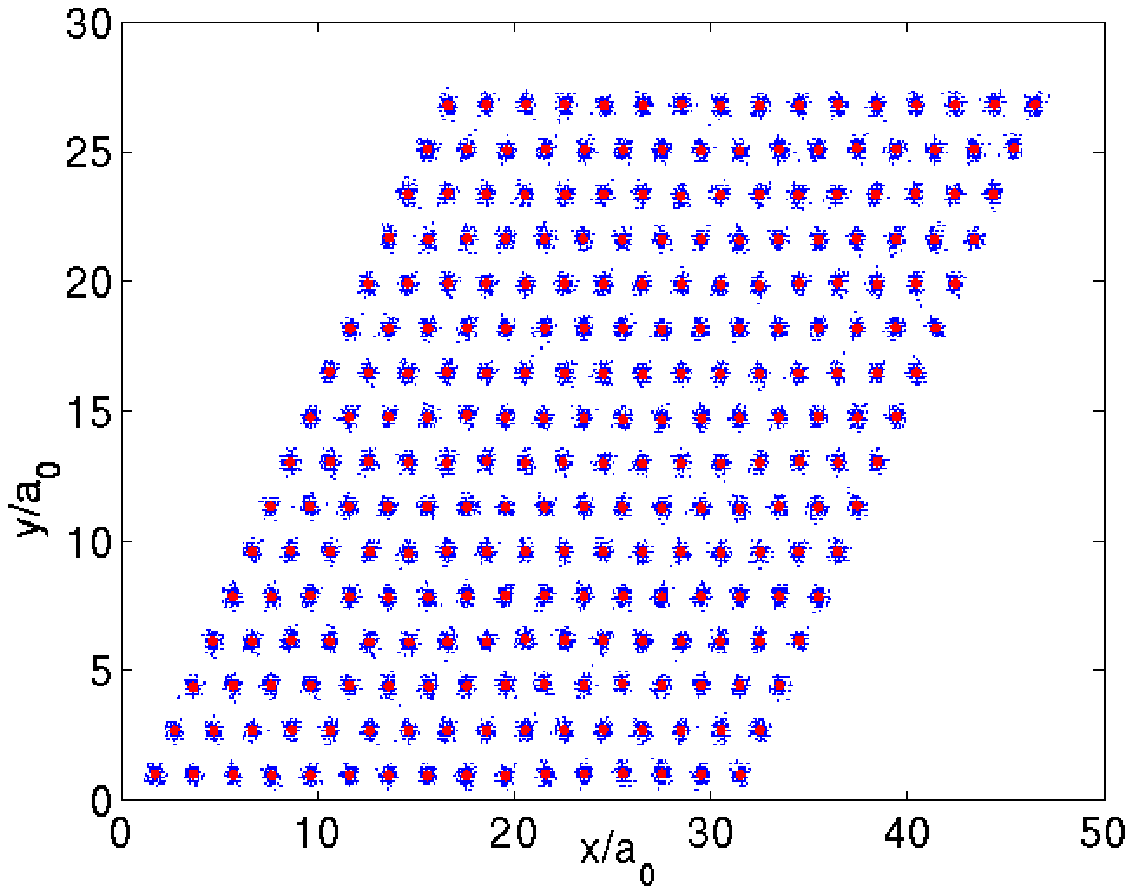}
\includegraphics [scale=0.4] {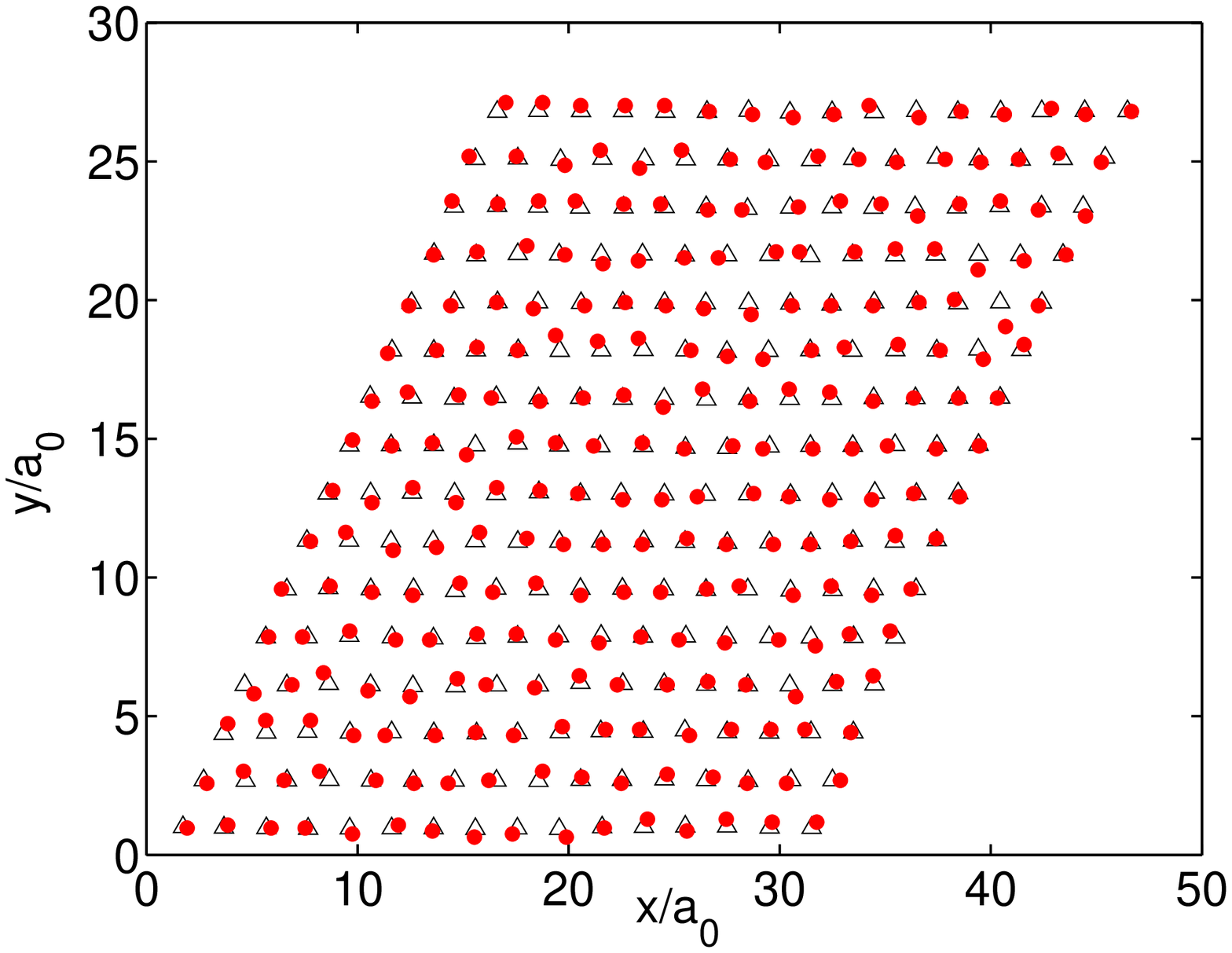}
\includegraphics [scale=0.4] {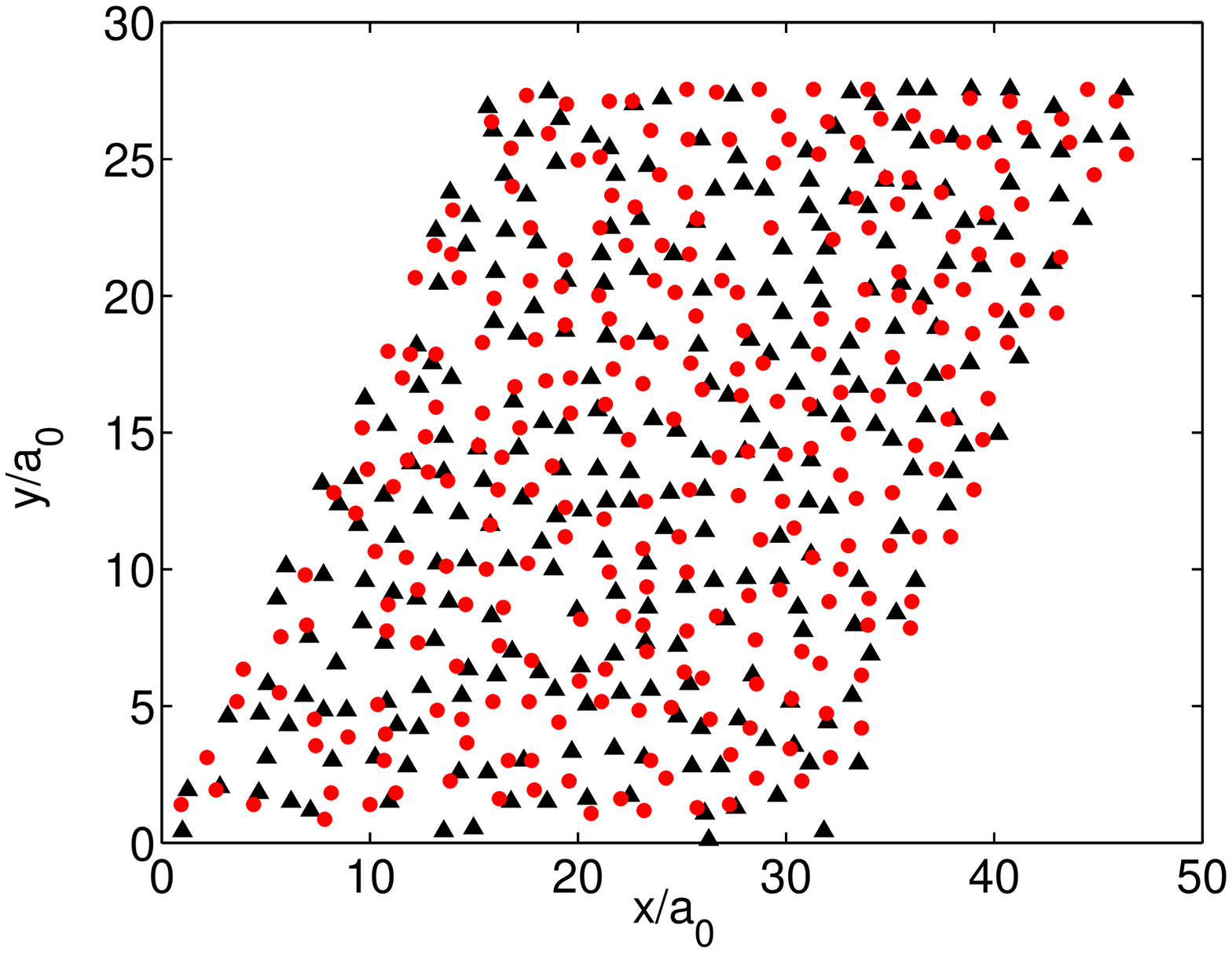}
\caption{\label{fig1} (Color online). Visualization of density structures
at different kinds of free energy minima. 
The first two panels correspond to an ordered minimum. In
the first, the (blue) dots indicate, for each of the 128 layers present,
the positions of pancake vortices in that layer. The (red) filled circles 
denote the
average positions of ``vortex lines'' obtained by connecting adjacent 
vortices on neighboring layers.  In the second panel, the 
(red) full circles have the same meaning
and the (black) triangles represent the vortex 
position in a randomly chosen layer.
In the third panel, the (red) full circles and the (black) triangles represent
the positions of pancake vortices on two adjacent layers in
a disordered minimum. In all cases $T=17.0 K$ and $s=10.0 K$.}
\end{figure}

To visualize the vortex lattice, the local peak densities are extracted from
the set of $\rho_{n,i}$ values by defining site $i$ in layer $n$ 
to be a peak location
when it corresponds to a local maximum of the $\rho_{n,i}$ such that the value
$\rho_{n,i}$ exceeds that of any $\rho_{n,j}$ in the same plane
and within a radius $a/2$ from $i$. The
positions of such ``vortex sites'' in each layer can then be plotted. 
An example is given in Fig.~\ref{fig1}.
The results there are all at $s=10.0 K$ and $T=17.0 K$. 
In the first
panel, a small (blue) dot denotes the position of a vortex site at any one
of the 128 layers.
The results plotted
correspond to a  state  obtained starting with ``crystal-like'' initial
conditions, as explained above. 
The vortex positions  are clearly
clustered into well-defined groups that form a triangular pattern. For
such minima, we can define ``vortex lines'' by joining the vortex positions
in each group on different layers. The (red) filled circles in the first
panel indicate the average positions of these vortex lines.
These average positions form a slightly distorted triangular lattice, with
considerable variation, however,  from layer to layer. 
The typical size of the clusters of (blue) dots represents the degree of
transverse layer-to-layer wandering of the vortex lines. 
In the second panel, the same point is emphasized by displaying again
the average positions as (red) filled circles, 
and the vortex positions at a randomly chosen
layer as triangles: the degree of crystalline
order in the average positions of the vortex lines is higher than that in
the vortex positions on a typical layer -- the random transverse displacements
of the vortex positions on different layers from the ideal triangular lattice
partly cancel out in the calculation of the average positions of the
vortex lines. From these results 
and other evidence discussed below, we identify
minima of this kind as representing the BrG phase.

If one plots the vortex positions as in the first panel of Fig.~\ref{fig1} 
for a minimum obtained from liquid-like
initial conditions at the same $T$ and $s$, one obtains 
a nearly uniform space-filling
plot: the positions of the 
pancake vortices on different
layers are nearly uncorrelated at this minimum.
This difference between the two kinds of local free energy minima 
is emphasized in the third panel of Fig.~\ref{fig1} 
where the vortex positions in two neighboring layers are plotted for a
disordered minimum obtained for the same pin configuration, 
$T$, and $s$ as those for the plots in the first two panels.
Disordered local minima of this kind are identified below as representing the 
VL phase.

\begin{figure}
\includegraphics [scale=0.8] {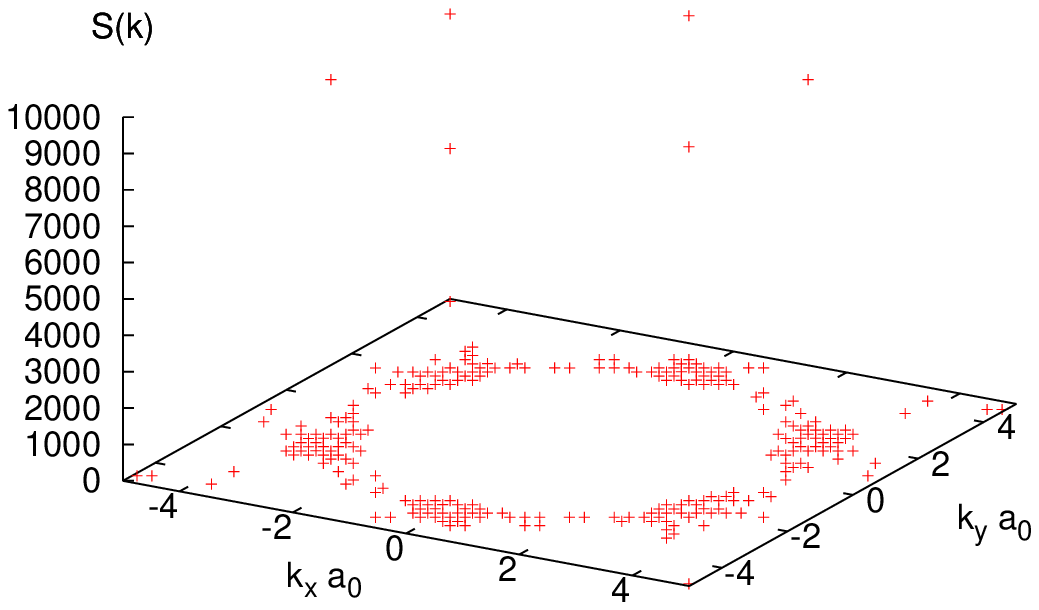}
\includegraphics [scale=0.8] {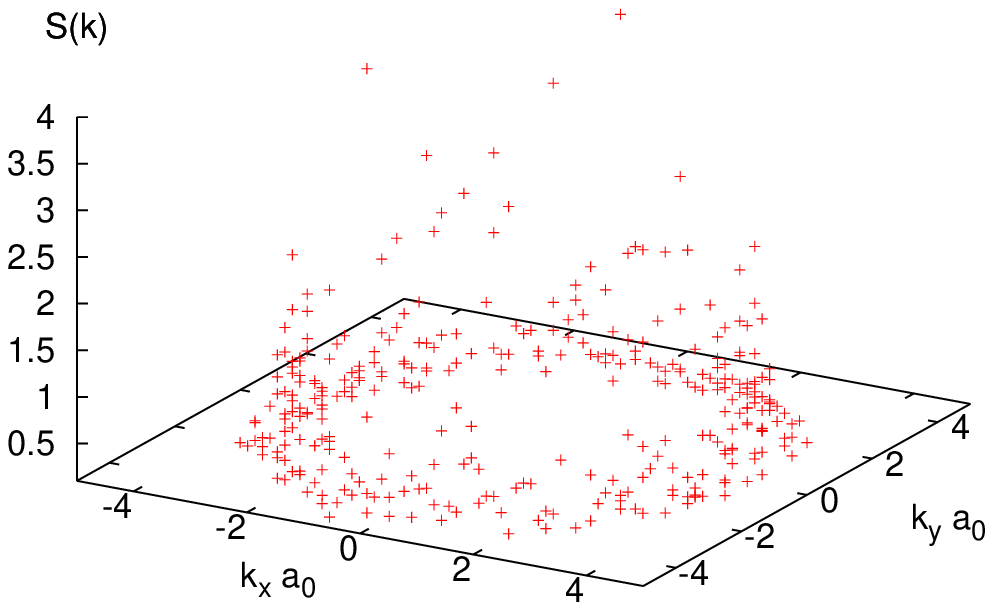}
\caption{\label{fig2} (color online).The static structure factor  as defined
in Eq.~(\ref{sofk}) plotted versus the two dimensional 
wavevector (in units of $a_0^{-1}$) 
  at $T=17.0 K$ and $s=10.0 K$ for the equilibrium ordered 
state  under these conditions (first
panel) and for the metastable disordered state
(second panel).}
\end{figure}

It is very useful to discuss also
the difference between ordered and disordered states
in terms of correlation functions. We do this  in the
next two figures. Consider first  Fig.~\ref{fig2}.
There  we plot the ``two-dimensional'' static
structure factor $S({\bf k})$ defined as:
\begin{equation}
S({\bf k})=|\rho({\bf k},k_z=0)|^2/(N_v N_L)
\label{sofk}
\end{equation}
where $\rho({\bf k},k_z)$ is the discrete Fourier 
transform of the  $\{\rho_{n,i}\}$ in terms of the wavevector $({\bf k},k_z)$
(the $z$-axis is normal to the layers).
In the figure $T=17.0 K$ and $s=10.0 K$. In the first panel
$S({\bf k})$ is plotted for the ordered state (which is, as we shall see, 
the equilibrium one in this case) while in the other panel the same
quantity is plotted for the disordered 
state obtained from uniform initial conditions,
which is locally stable at these values of $s$ and $T$. The difference
between the two states can be readily seen: note the large difference between
the vertical scales in the two plots and the Bragg-like peaks in the
ordered case.

\begin{figure}
\includegraphics [scale=0.7] {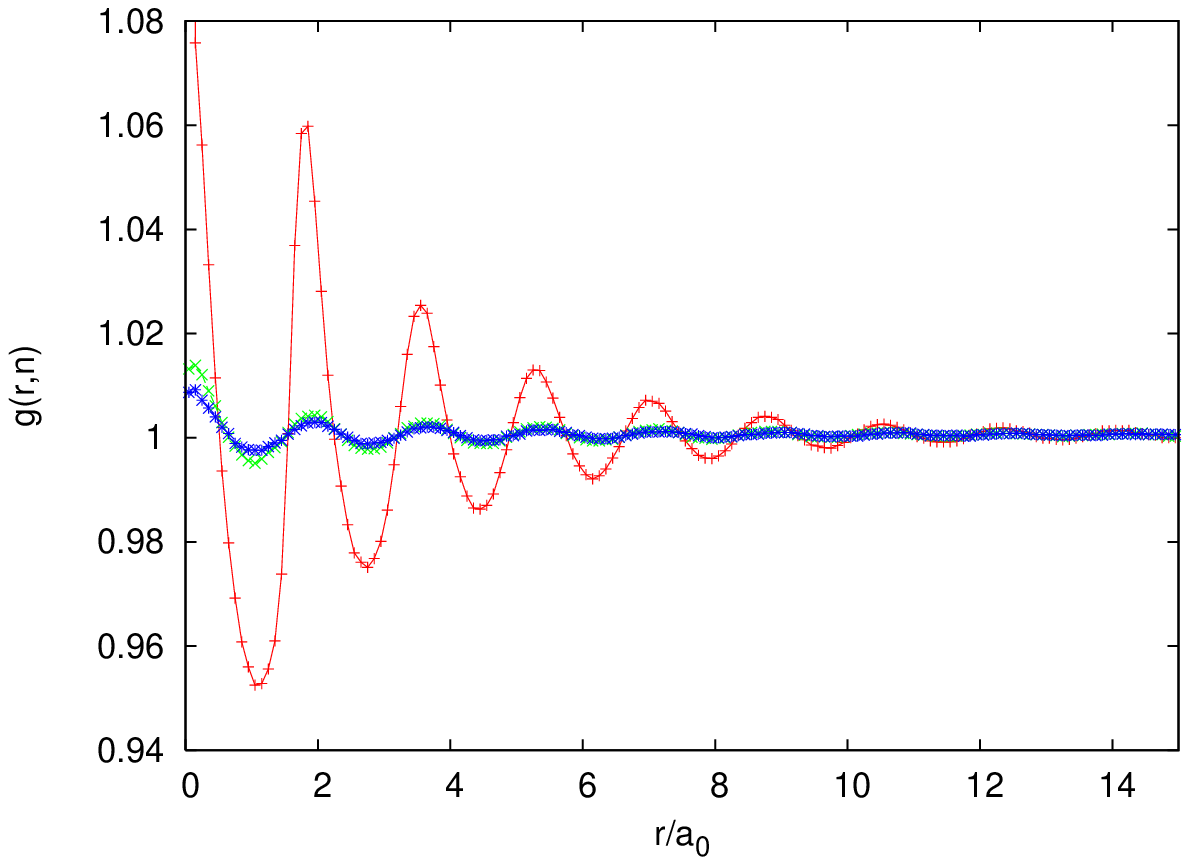}
\includegraphics [scale=0.7] {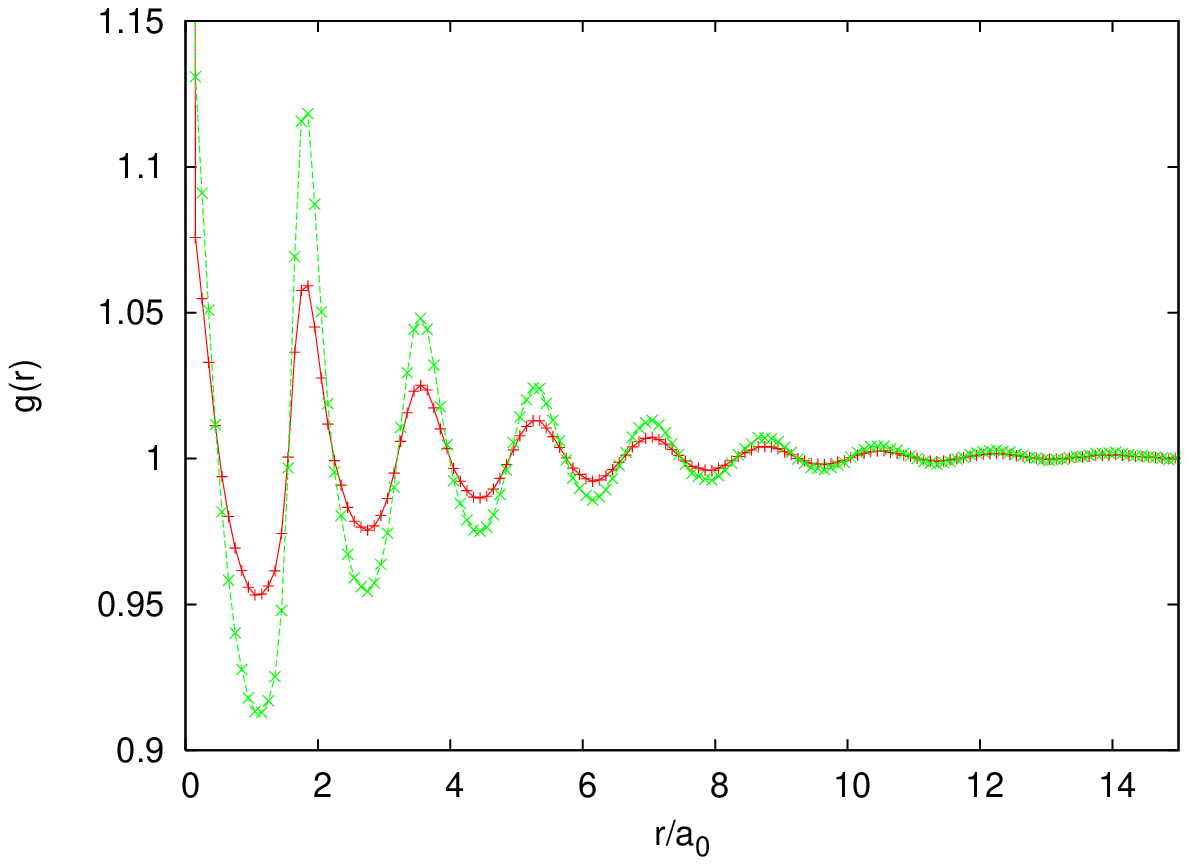}
\includegraphics [scale=0.7] {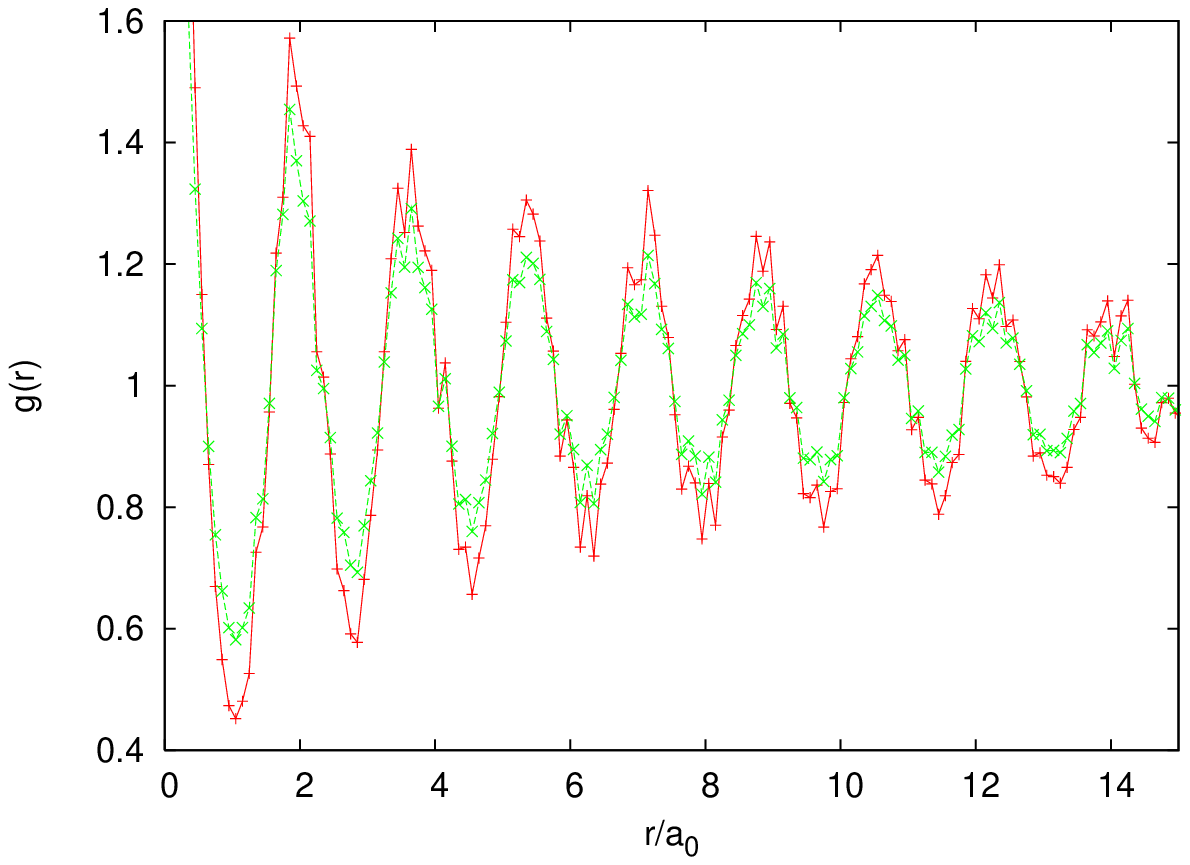}
\caption{\label{fig3} (Color online) Real-space angularly averaged
correlation functions (see text)
plotted versus two-dimensional dimensionless in-plane
distance at $T=17.0 K$. The first panel displays (top to
bottom curves at small $r$) $g(r,n)$ for $n=0$ (red), $n=1$ (green),
and $n=10$ (blue)
at $s=10.0 K$, in the disordered state. The second and third panels
show $g(r) \equiv g(r,n=0)$ for the disordered and ordered states, respectively,
at $s=10.0 K$ and $s=12.0 K$, with the paler (green) curve corresponding
to the higher $s$.
}
\end{figure}

Alternatively, one can look at real-space correlation functions obtained from
the discrete Fourier transform of $S({\bf k},k_z)$. We define $g(r,n)$
as the angularly averaged correlation of the time-averaged local densities
at two points separated by $n$ layers and in-plane distance $r$. 
The same-plane correlation function is $g(r) \equiv g(r,n=0)$.
For $n \neq 0$,  $g(r,n)$ represents
correlations between the vortex densities in different planes. 
We normalize $g(r,n)$ by $\rho_0^2$,
so that it approaches unity at large values of $r$ for all $n$. 
These correlation functions
are {\it different} from the pair distribution function ${g^\prime}_n(r)$
shown in Fig.~\ref{fig0}: the $g(r,n)$ 
represent correlations of time-averaged local densities, whereas
${g^\prime}_n(r)$ are the equal-time correlation functions of the instantaneous
local density. In particular, $g(r,n)$ equals exactly  unity for all
$n$ and $r$ for a homogeneous vortex liquid in the absence of any
pinning, while ${g^\prime}_n(r)$ for a homogeneous liquid shows oscillations 
(seen in Fig.~\ref{fig0}) that reflect the 
short-range correlations present in a liquid. In the presence of a pinning
potential that makes the time-averaged local density in the liquid
phase inhomogeneous,
the correlation function $g(r,n)$ represents the spatial correlations of this
pinning-induced inhomogeneity. 

Results  
for $g(r,n)$ are presented in Fig.~\ref{fig3}.
In the first panel $g(r)$, $g(r,1)$ and
$g(r,10)$ 
at $T=17.0$ and $s=10.0 K$ for the liquid-like
metastable state are shown.  The decay of the correlations with $n$ is 
clear: the vortex densities on different layers are only weakly
correlated in the liquid-like phase, in agreement with the results shown
in the third panel of Fig.~\ref{fig1}. The corresponding
plot for the ordered state at the same values
of $T$ and $s$ (not displayed) would show  no decay with $n$
except at very small values of $n$,  
indicating that the vortices on different
layers are close to registry in this state.
 In particular, $g(r=0,n)$
is  found to be 
essentially
independent of $n$ for $3 < n \lesssim N_L/2$ in the ordered
phase, showing absolutely no sign of decay for large values of $n$. 
 In the other two
panels, the variation of $g(r)$ with $s$ is shown for both the 
disordered and the ordered states and two values of $s$. The two cases
are again in stark contrast. For the disordered, 
liquid-like minimum (second panel), the
amplitude of the damped oscillations of $g(r)$ is much smaller than that for
the ordered, crystal-like minimum (third panel). 
The amplitude of these oscillations
increases with $s$ for the liquid-like minimum because the pinning-induced
inhomogeneity in the local density increases as the pinning gets stronger,
whereas for $g(r)$ in the ordered 
minimum, this amplitude decreases as $s$ is increased, reflecting  that 
stronger pinning leads to larger distortions of the crystalline structure.
One could similarly plot the variation
of $g(r)$ with temperature: in this case the peak heights decrease as $T$
increases. All these features of $g(r,n)$ for the liquid-like state are very
similar to those found in an analytic study~\cite{gautamprl} of the effects
of pinning disorder on the correlations of a layered vortex liquid.

\begin{figure}
\includegraphics [scale=0.7] {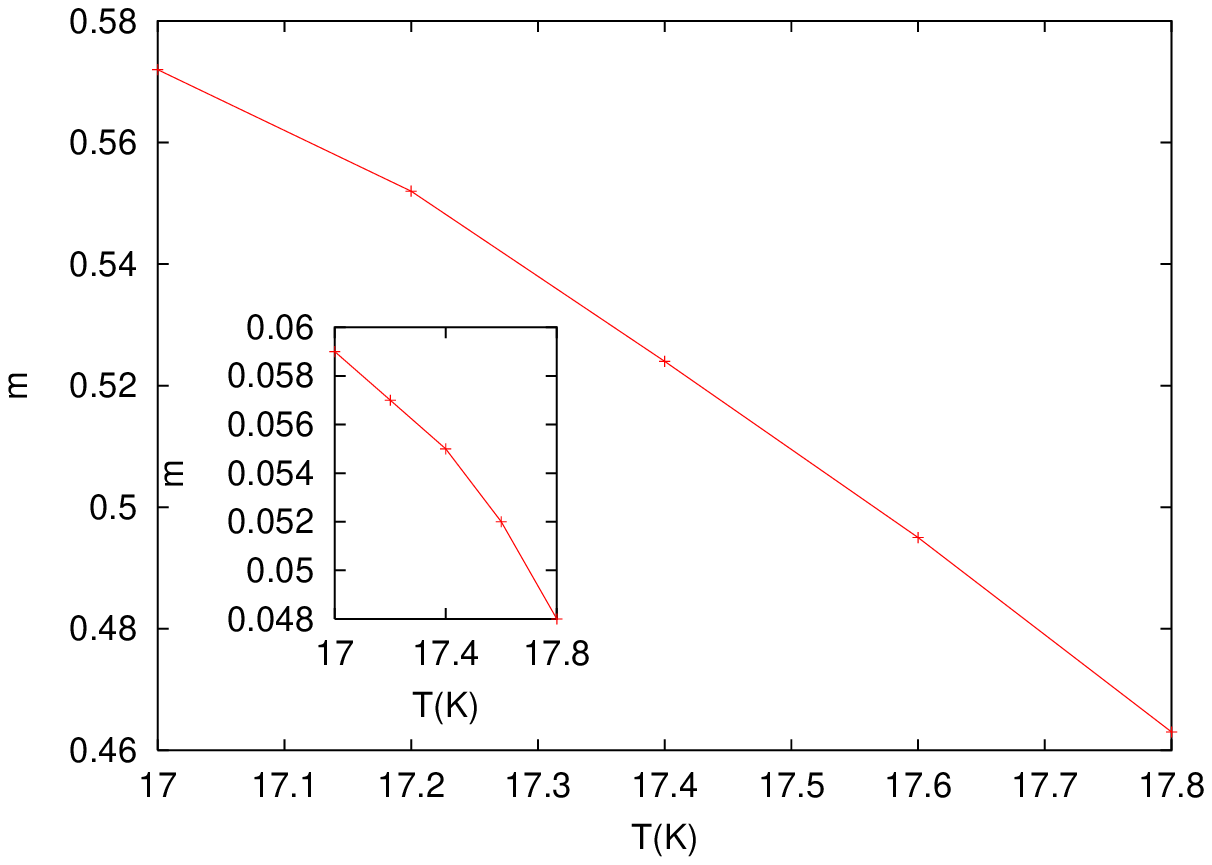}
\includegraphics [scale=0.7] {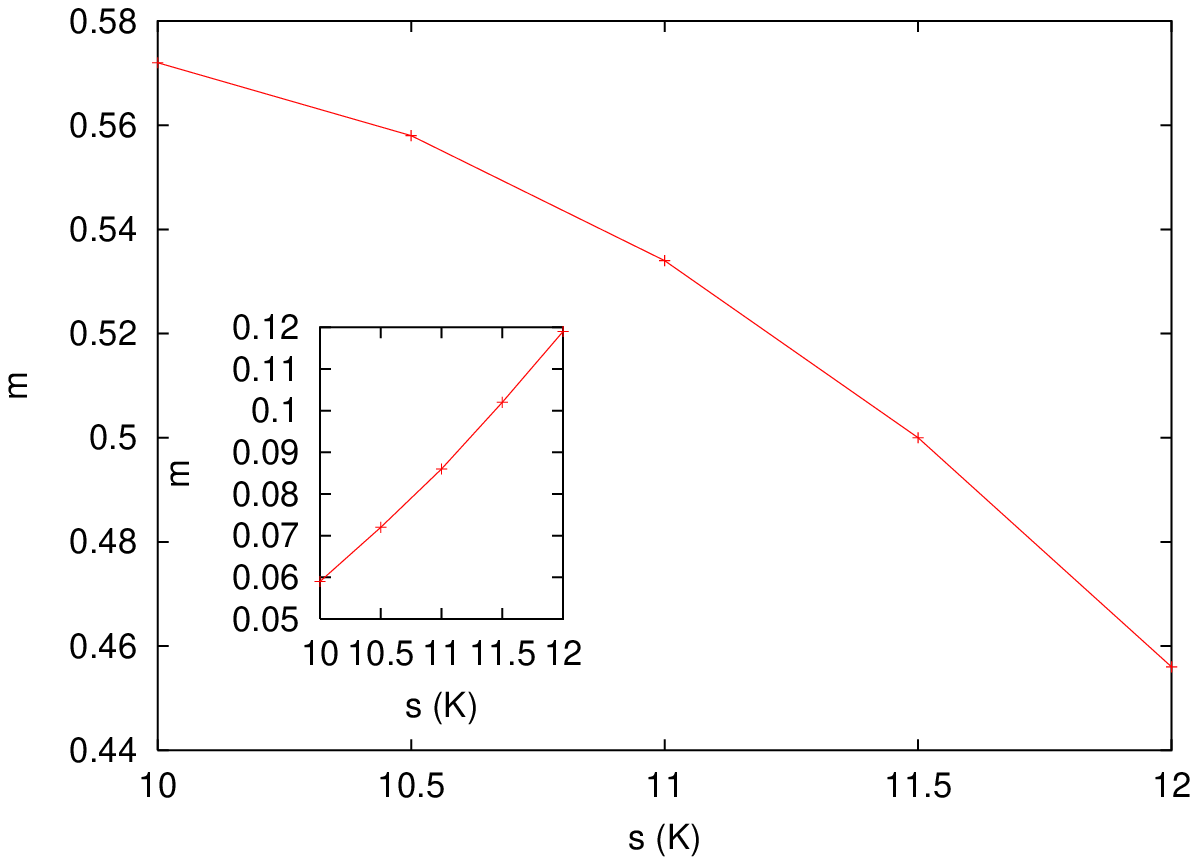}
\caption{ \label{fig4} (Color online).First panel: The order parameter $m$, 
as defined in the text, plotted versus
temperature at constant $s=10.0 K$ for the ordered state (main plot) and
the disordered state (inset). Second
panel: The same quantity plotted versus
pin strength $s$ at constant $T=17.0 K$ for the ordered state (main plot) and
the disordered state (inset). The solid lines are guides to the eye.}
\end {figure}

From the correlation functions, an appropriate order parameter can
be extracted. A convenient definition is $m \equiv g(r_m)-1$ where $r_m$
is the first nonzero value of $r$ for which $g(r)$ has a peak. An
alternative definition~\cite{prbd} in terms of $S({\bf k},k_z=0)$ yields 
very similar results.
In the first panel of Fig.~\ref{fig4}, this quantity is plotted 
as a function of the temperature 
at constant $s=10.0K$ for both the ordered
and (inset) the disordered state. It is, as it should be, much larger in the
former case and it decreases with increasing $T$ 
for both states. As we shall see below, the transition
temperature  at this value of $s$ is near $T=17.44 K$. In the 
second panel of Fig.~\ref{fig4},
$m$ is plotted versus $s$ at constant $T=17.0K$ for the same two cases. The
transition as $s$ is increased at this $T$ occurs at $s=11.1 K$. We see
that $m$ now {\it increases} with $s$ in the disordered state while
remaining much smaller than its values in the ordered state. In the ordered
state, $m$ decreases with $s$, as expected. 

\begin{figure}
\includegraphics [scale=0.33] {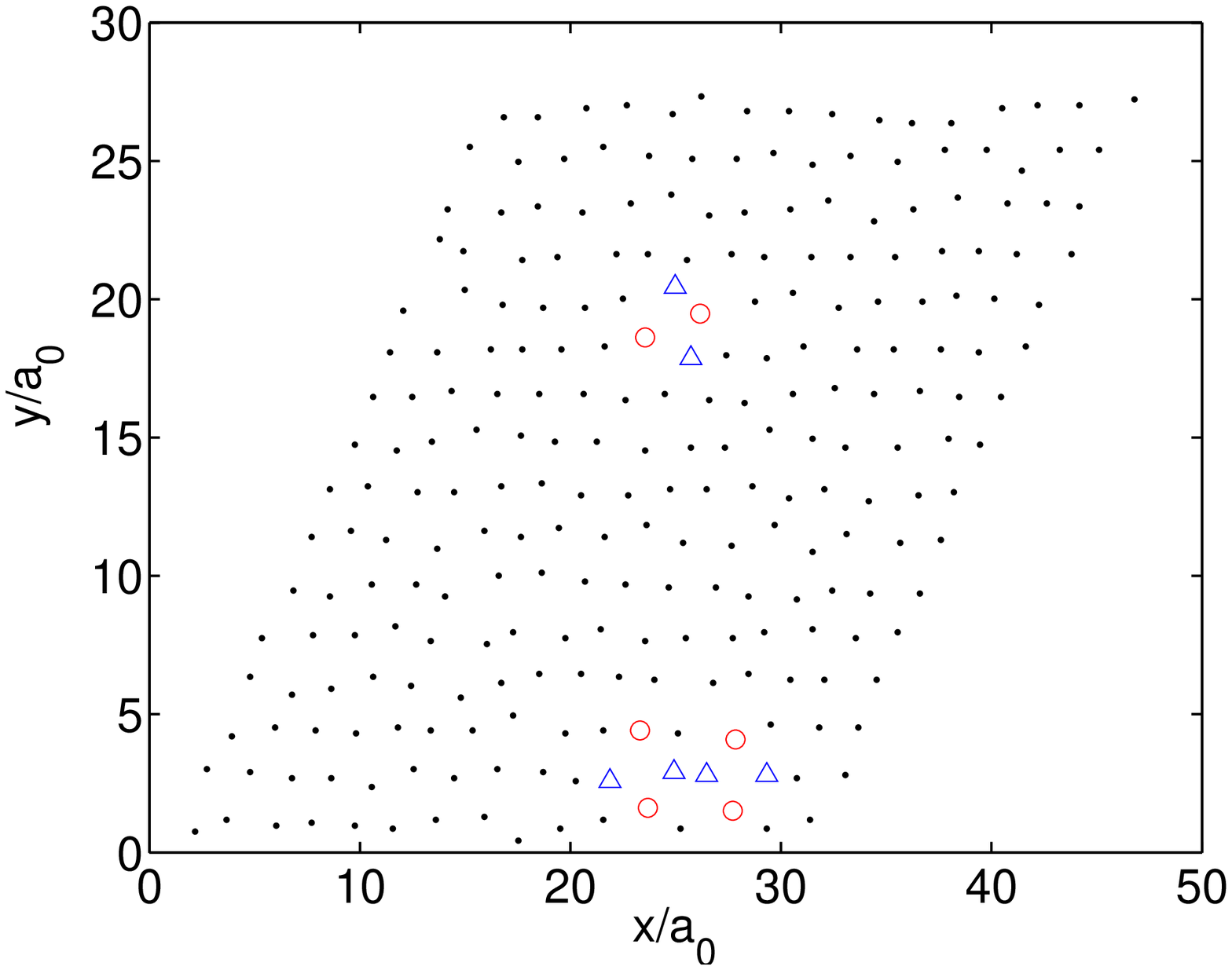}
\includegraphics [scale=0.33] {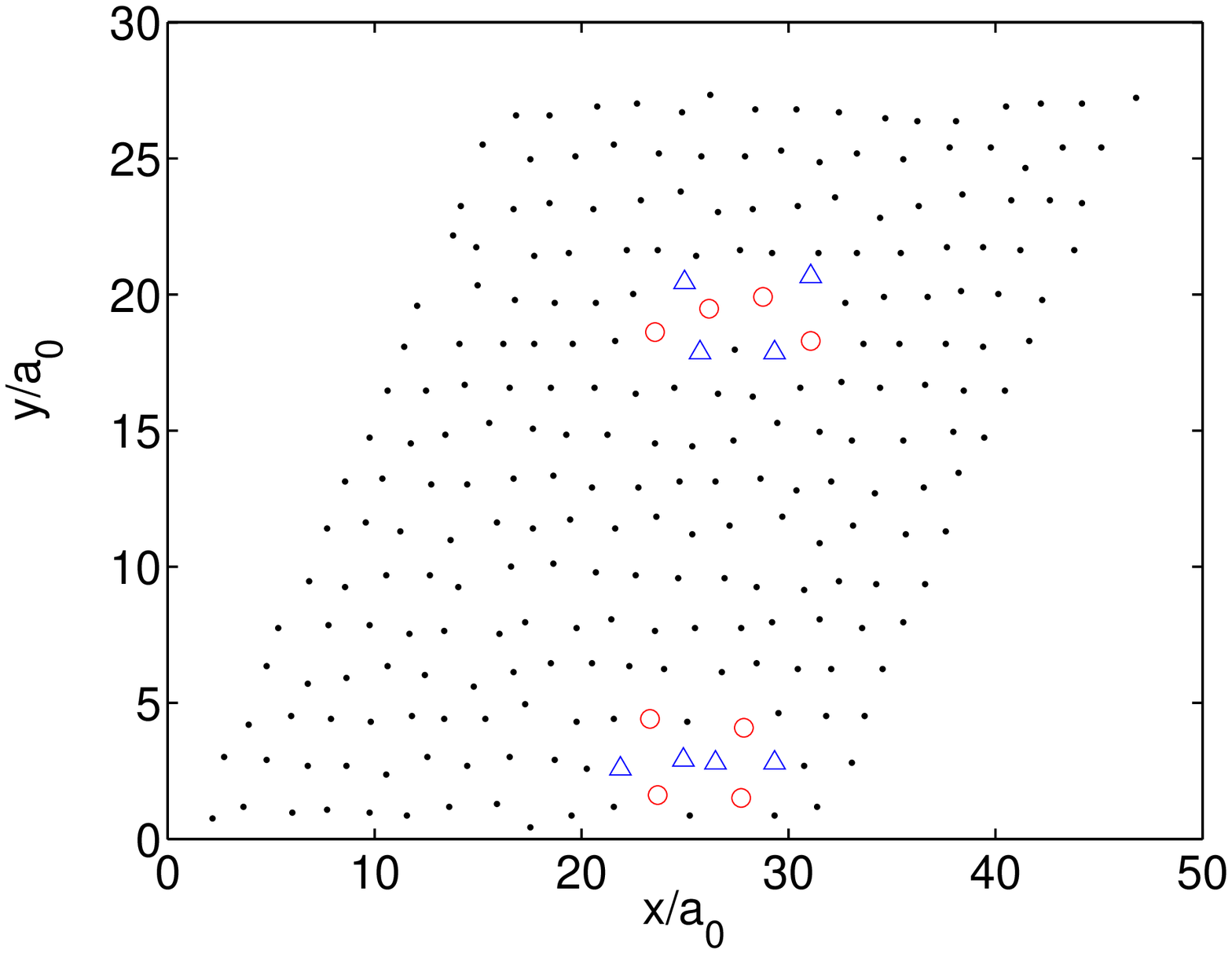}
\includegraphics [scale=0.33] {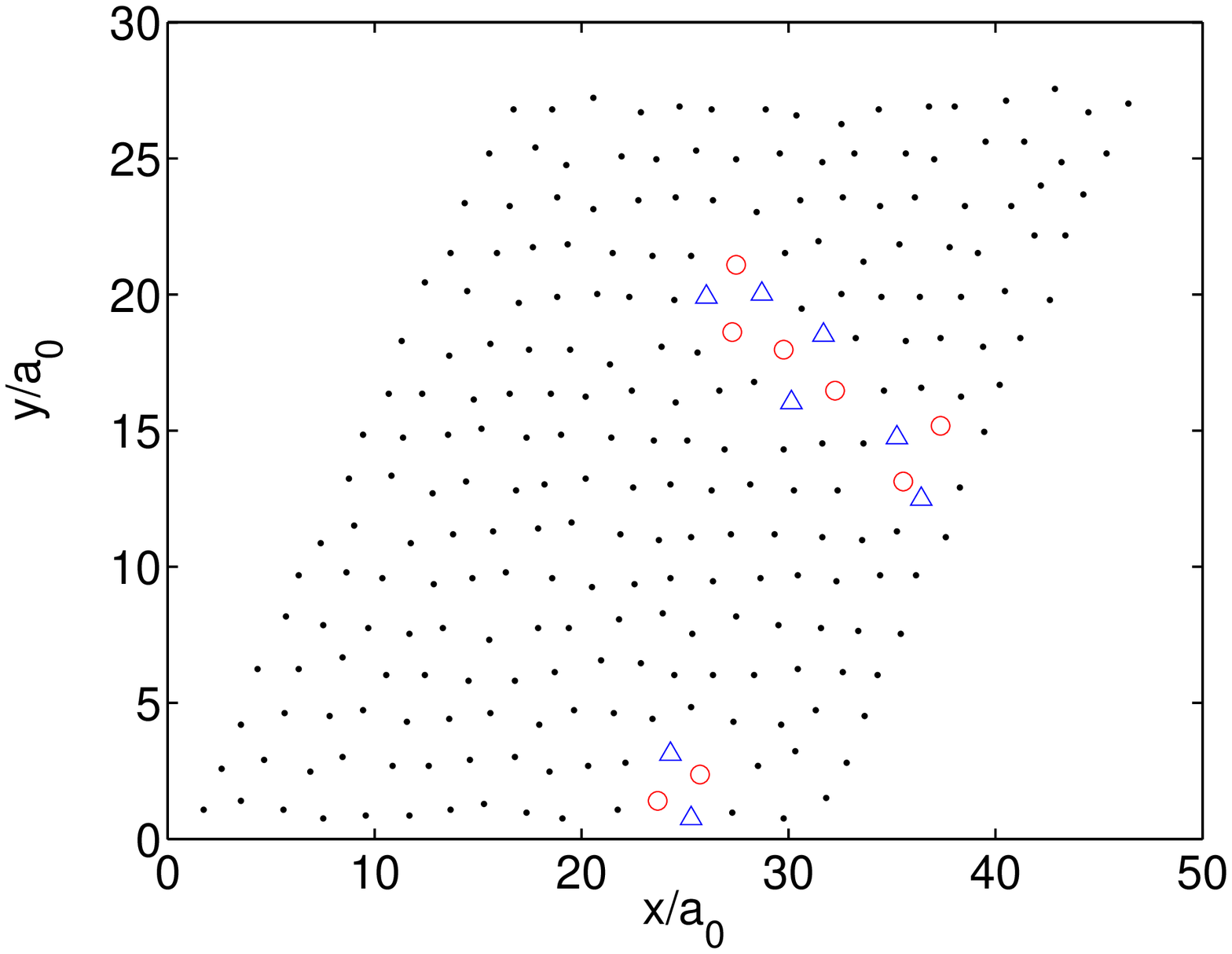}
\includegraphics [scale=0.33] {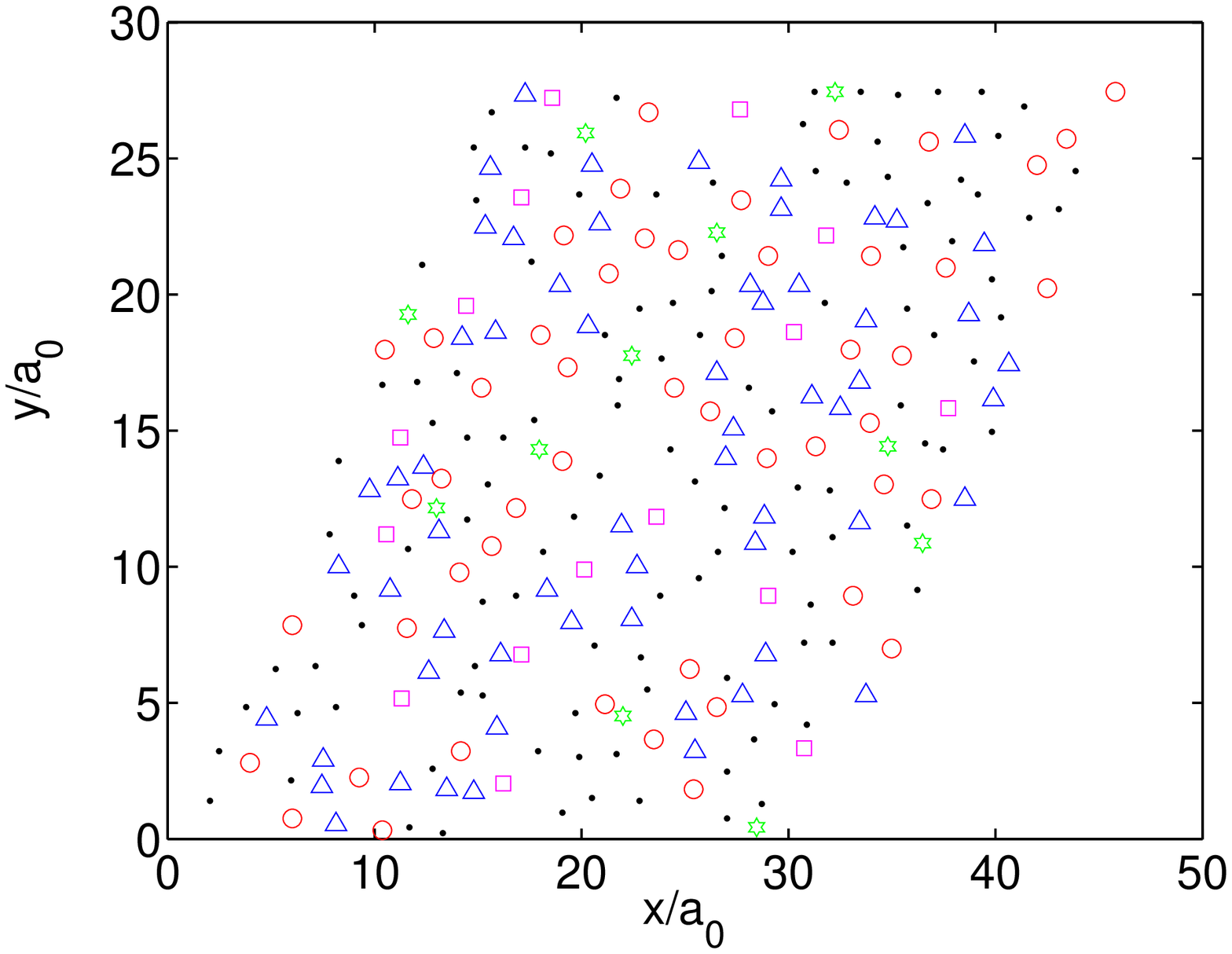}
\caption{ \label{fig5} (Color online)
Voronoi plots of the vortex lattice. The (black) dots denote ordinary sites
with six neighbors, the (red) circles sites with seven neighbors and the
(blue) triangles sites with five neighbors. The first three panels are
for the ordered state with $s=10.0 K$,
$T = 17.0 K$ (top left panel), $s= 10.0 K$, $T=17.4 K$ (top right panel),
and $s = 12.0 K$, $T=17.0K$ (bottom left panel).
The last panel is for
the disordered state at $s=10.0 K$, $T = 17.6 K$. The (green) filled circles
and (magenta) squares in this plot represent sites with eight and four
neighbors, respectively.}
\end {figure}

It is also useful to visualize the defect structure of the phases by means
of Voronoi plots of the vortex lattice. In these plots the number of neighbors
of each vortex lattice site 
in a given layer is found via a Wigner-Seitz construction.
One then plots with different symbols the lattice points having six
neighbors (this would be all the sites for a defect-free lattice) and those
having e.g. seven or five neighbors (which correspond to single disclinations:
neighboring pairs of disclinations of opposite sign correspond to
dislocations). The results are shown in Fig.~\ref{fig5}.
The first three panels are for the ordered state, at $s=10.0 K$,
$T = 17.0 K$ (top left panel), $s= 10.0 K$, $T=17.4 K$ (top right panel),
and $s = 12.0 K$, $T=17.0K$ (bottom left panel). The system is deep in
the ordered phase for the first set of values of $s$, $T$, while the other
two sets represent points close to the phase boundary (see below)
between the ordered and disordered phases. In each case, the defect
structure in the layer with the largest number of defects is plotted. One can
see a few dislocations in each of the three plots, but these dislocations occur
in tightly bound clusters with zero net Burgers vector. 
The number of defects is small and increases only weakly
with $T$ for fixed $s$, and with $s$ if $T$ is fixed. 
One clearly has a rather well ordered crystal-like structure even for values
of $s$ and $T$ close to the phase boundary. A feature to note in these
plots is that all these structures are
single-domain: domain boundaries which appear in Voronoi plots\cite{prbd}
as lines of defect pairs are clearly absent. 
By contrast, the last panel in this figure, which 
depicts the
disordered state obtained for $s=10.0 K$ and $T = 17.6 K$ (the disordered state
is the equilibrium one for these parameter values, see below), shows 
a very large number of randomly distributed defects. 
The structure of this state may be classified as amorphous -- there is no
indication of the presence of grain boundaries separating crystalline 
regions with different orientations.

\begin{figure}
\includegraphics [scale=0.4] {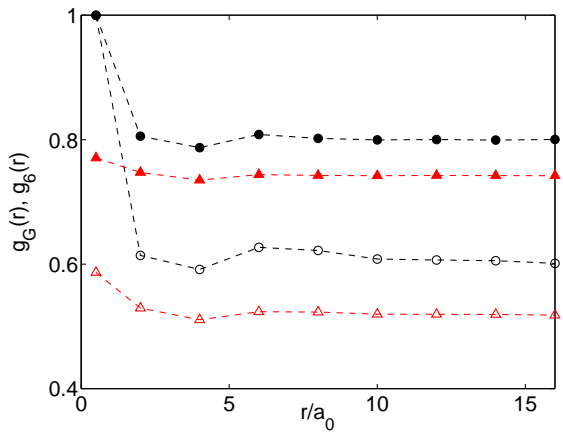}
\caption {\label{fig6} (Color online). The bond-orientational
correlation function
$g_6(r)$ ((red) triangles) and the translational correlation 
function $g_G(r)$ ((black) circles) as defined in the text, plotted as
functions of in-plane distance $r$ for the ordered state at $s=5.0 K$,
$T=17.0 K$ (filled symbols), and $s=10.0 K$, $T = 17.0 K$ (empty symbols). The
dashed lines are guides to the eye.}
\end{figure}

The degree of orientational order in vortex 
configurations is most conveniently defined
in terms of the bond-orientational correlation function $g_6(r)$ defined
as:
\begin{subequations}
\label{angular}
\begin{equation}
g_6(r)=\langle\psi(\bf{r})\psi(0)\rangle
\label{ang1}
\end{equation}
where, as before, ${\bf r}$ is a vector in the layer plane, the brackets
$\langle\cdots\rangle$ denote an average over the choice
of the origin in a particular layer, over all the layers
in the computational sample,
 and over angles.
The field $\psi({\bf r})$ is:
\begin{equation}
\psi({\bf r})=\frac{1}{n_n} \sum_{j=1}^{n_n} \exp[6i\theta_j({\bf r})]
\end{equation}
\end{subequations}
where $\theta_j({\bf r})$ is the angle made by the bond connecting
a vortex at ${\bf r}$ to its $j$-th in-plane neighbor and a fixed
axis, and $n_n$ is the number of in-plane 
neighbors of the vortex at ${\bf r}$, as
determined from a Voronoi construction. 
Similarly, one can  define 
a ``translational correlation function'' $g_G(r)$ of the vortex lattice 
by an equation identical
to the right-hand side of Eq.~(\ref{ang1}), but with the field $\psi$
replaced by
\begin{equation} 
\psi_G({\bf r})=\exp(i {\bf G} \cdot {\bf r}).
\label{bo}
\end{equation}
Here ${\bf G}$ is one of the shortest nonzero two-dimensional
reciprocal lattice vectors of the triangular vortex lattice
in the absence of pinning. We average over the  six equivalent $\bf G$'s
and over all the layers. 

Examples of these correlation functions are plotted in Fig.~\ref{fig6} where
results are shown for the ordered minima at $T=17.0 K$ and two different
values, $5.0 K$ and $10.0 K$, of the pinning strength $s$. The function
$g_6(r)$ is shown by the (red) triangles and $g_G(r)$  by the (black)
circles. In each case, the upper plot corresponds to the smaller value of $s$.
Both these functions appear to saturate to constant values at large $r$,
indicating the presence of long-range bond-orientational order and (nearly)
long-range translational order (the length scales considered here are 
probably too short to show the power-law decay of $g_G(r)$ 
expected~\cite{natt,giamarchi} for large
$r$ in a BrG phase) for both values of $s$. From these results, and those 
shown in Fig.~\ref{fig1} (top two panels), Fig.~\ref{fig2} (top panel),
Fig.~\ref{fig3} (bottom panel), and Fig.~\ref{fig5} (top three panels), we
conclude that the ordered minima may be identified as representing the BrG
phase.

\begin{figure}
\includegraphics [scale=0.7] {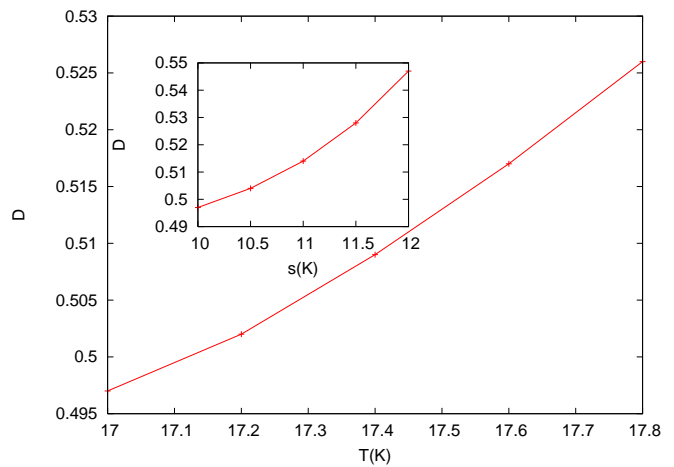}
\caption{\label{fig7} (Color online). The average width $D$ of the local density
peaks 
in the ordered phase (as defined in the text), 
measured in units of $a_0$. In the main
plot, results are shown versus $T$ at pin strength $s=10.K$ while
in the inset $D$ is  plotted versus $s$ at $T=17.0$}
\end{figure}

The
 width of the 
local density peaks of the vortex lattice at
free-energy minima quantifies the degree of
thermal and disorder-induced wandering of the pancake vortices from their
time-averaged positions. We have studied how the average width of the local
density peaks depends on $T$ and $s$. The width 
$D_{n,i}$ of a local density peak at computational mesh point $i$ on layer $n$
is calculated from the relation
\begin{equation}
D^2_{n,i} \equiv \frac{\sum^\prime_j r_{ij}^2 \rho_{n,j}}{\sum^\prime_j 
\rho_{n,j}}, \label{width}
\end{equation}
where 
$r_{ij}$
is the distance between mesh points $i$ and $j$ on the same layer, 
and the sum $\sum^\prime_j$
is over the mesh points on the same layer that lie inside a unit cell of the
vortex lattice centered at mesh point $i$.
A similar calculation was carried out in our earlier
study~\cite{prbv} of the Lindemann parameter at the melting transition of the
vortex lattice in the absence of pinning. Here, the widths of
different local density peaks are not all the same, and we define the average
width $D$ as the average of $D_{n,i}$ over all the local density peaks in 
the sample. The variation of $D$ with $T$ and $s$ is shown
in Fig.~\ref{fig7} where we plot $D$ in units of $a_0$ versus 
$T$ at constant $s=10.0 K$ in the main plot and at constant $T=17.0 K$
versus $s$ in the inset. 
As expected, $D$ increases with $T$ for fixed $s$. 
The value of $D$ at a fixed $T$
increases with $s$ -- the local density peaks become more distorted and 
broader as the pinning strength is increased.
Melting of the ordered state (see next subsection) occurs when 
$D$ reaches a value
of approximately 0.51 (this is true in all cases, not just those shown
in the figure). Using the value of the equilibrium lattice constant, 
$a=1.988 a_0$, for  $B=0.2T$, we get the value of the
Lindemann parameter, ${\mathcal{L}} \equiv D/a$, to be approximately 0.26.
This is close to the value obtained for the system without pinning~\cite{prbv},
indicating that the Lindemann parameter at  melting
is nearly independent of the disorder strength $s$.

The root-mean-square (rms)
lateral displacement of a pancake vortex from the 
average position of the {\it vortex line} to which it belongs contains another
contribution, arising from the pinning-induced transverse wandering of the
position of the local density peak that represents the average position
of the pancake vortex. 
As mentioned in
connection with Fig.~\ref{fig3},  in
the ordered phase the local density peaks on different layers are  
approximately in registry. However plots such as those in 
Fig.~\ref{fig1} indicate that there is an appreciable variation in
the position of a vortex line as one moves from layer to layer -- the typical
width of the clusters of (blue) dots in the first panel of Fig.~\ref{fig1}
provides a measure of this variation. We have calculated the $T$- and
$s$-dependence of $D^\prime$, the rms displacement of the average positions of
the pancake vortices that form a vortex line from the average position of
the vortex line, further averaged over all vortex lines in an ordered state.
The $T$-dependence of this quantity is weak in the range of
temperatures considered -- it increases slightly as $T$ is increased. As
expected, $D^\prime$ increases as $s$ is increased at fixed $T$: its value in
units of $a_0$
at $T=17.0 K$ is 
of course zero at $s=0$, 
0.16 for
$s=5.0 K$, 0.23 for $s=7.5 K$, 0.27 for $s=10.0 K$, and 0.31 for $s= 12.0 K$.
Similar results are obtained at other temperatures. Thus, $D^\prime$
increases almost linearly with $s$ for relatively small values of $s$ and the
rate of increase becomes smaller as $s$ is increased further. 
$D^\prime$ is typically substantially smaller than $D$ for
the values of $s$ and $T$ studied. If the layer-to-layer random
displacements of the average positions of the pancake vortices from the average
position of the vortex line to which they belong and
the lateral displacements of the pancake vortices from their average
positions are assumed to be statistically independent, 
then the rms displacement of individual vortices from the avenge
position of the vortex line would be given by $D_{net} = (D^2 + D^{\prime 2})^
{1/2}$. This would lead to values of $D_{net}$ that are $\sim$ 15\%
higher than those of $D$ for the largest value of $s$ ($=12.0 K$) considered
here. As discussed in the next subsections, these results are important for
assessing the validity of the Lindemann criterion for the melting 
transition of the BrG state. 
 
We end this subsection with a few comments on the structure of the disordered
VL minima. As seen above, the arrangement of the vortices at these minima
may be classified as amorphous, not polycrystalline. While early decoration
experiments~\cite{deco1} on layered high-$T_c$ superconductors showed evidence
for an amorphous structure of the disordered phase, 
more recent experiments~\cite{deco2}
on ${\rm NbSe_2}$ have shown the occurrence of polycrystalline 
disordered structures. Also, simulations~\cite{sim1,sim2} of the structure of
two-dimensional vortex systems in the presence of random point pinning have
found polycrystalline structures. Our recent 
work~\cite{us1,us2,prbd} on layered superconductors with columnar pins
perpendicular to the layers also showed  polycrystalline
disordered states. It is therefore pertinent to inquire why polycrystalline
structures are not found in the present calculation. While a complete 
answer to this question is not yet available, we believe that this is
a consequence of the nature of the pinning potential considered here.
In a sample with a small concentration of columnar pins perpendicular
to the layers, the pinning potential (which is the same
in all layers) is large at a few points and zero in the
other parts of the sample. This favors the formation of 
polycrystalline structures: 
crystallites can form in the large 
``interstitial'' regions where the pinning potential is
zero, and since such regions on different layers are in registry, there is
no conflict between the preferred arrangements of vortices on different layers.
In contrast, the formation of local crystalline regions is less likely in the
situation considered here because the pinning potential is
nonzero at almost every computational mesh point. 
Also, since the pinning potentials
on different layers are now completely uncorrelated, the
formation of crystallites would be energetically favorable (if at all) 
in different regions
on different layers, and such regions would, in general, not be in registry. 
So, the interlayer interaction that favors vortices on adjacent layers being
in registry would act against the formation of crystalline patches. The last
factor is, of course, absent in two-dimensional systems and less important for
surface layers probed in decoration experiments. This may explain
the observation of polycrystalline structures in 
two-dimensional simulations and some decoration experiments.

\subsection{Phase diagram}

We have seen then that there are two kinds of local minima 
of the free energy in the $(s,T)$ region we have studied.
The ordered minimum is to be identified as a Bragg glass (BrG) while the
disordered one is a vortex liquid (VL, whether it should
be called vortex glass is discussed below).

\begin{figure}
\includegraphics [scale=0.6] {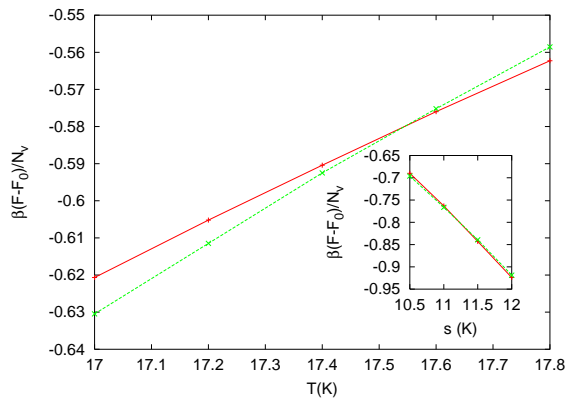}
\caption {\label{fig8} (Color online). Examples of the free energy
per vortex, in units of $k_BT$, plotted versus $T$ at constant $s=10.0 K$ (main
plot) and versus $s$ at constant $T=17.0 K$ (inset). The (red) plus signs are
data for the BrG phase and the (green) crosses for the disordered phase.
The data points are connected by  straight lines. Statistical error resulting
from pin configuration average would result in error bars smaller than
the symbols. First-order transitions occur at the crossings.}
\end{figure}  

Where both kinds of minima are locally stable, the equilibrium state is found
by comparing their free energies, which is trivial in our 
procedure. The transitions then are found by looking for free energy
crossings. Examples are shown in Fig.~\ref{fig8}, where results 
are shown as functions
of $T$ and $s$. We find that in all cases the transition is of first order:
as in Fig.~\ref{fig8}, the derivative of the free energy at the transition
is discontinuous.
In general, the BrG phase becomes unstable slightly above the transition: 
if one warms
up that phase (or increases $s$) beyond the melting point one converges, after a
large number of iterations, to the disordered phase. The reverse is not true, it
is possible to substantially supercool the VL phase:
the VL minimum remains locally stable in the whole $(s,T)$ region 
considered here ($0 \le s \le 12.5 K,\, 16.0K \le T \le 
21.0K$). 
This is consistent with experimental results~\cite{andrei} 
that suggest that
the BrG phase in ${\rm NbSe_2}$ exhibits a spinodal 
instability as the temperature is increased above the transition point, 
whereas there is no limit of metastability of the high-temperature
disordered state as it is cooled to temperatures lower than the transition
temperature. This is different from the behavior found in our recent 
study~\cite{prbd,us1,us2} of the vortex system in the presence of a small
concentration of random columnar pins -- in that case, the high-temperature VL 
minimum becomes unstable as the temperature is decreased below the transition
point. The absence of a spinodal instability of the VL minimum in the present
case is probably related to the fact (discussed above) that the formation of 
large crystallites which would make the VL minimum unstable is unlikely.

\begin{figure}
\includegraphics [scale=0.6] {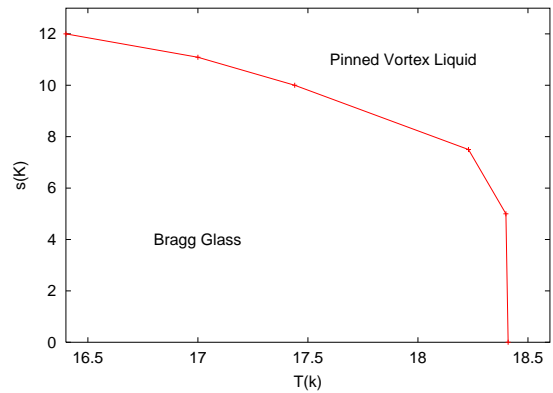}
\caption{\label{fig9} (Color online). Phase diagram in the $s-T$ plane at constant field
$B=0.2 T$. The symbols denote data points for the free energy
crossings where first-order
phase transitions between BrG and VL phases 
are found. They are connected by solid segments
which thus represent the locus of first-order transitions.}
\end{figure}

From free energy results such as those shown in Fig.~\ref{fig8} for
a variety of values of $s$ and $T$, it is possible to map the phase diagram
in the $s-T$ plane. The results, averaged over all samples and at constant
$B=0.2T$, are shown in Fig.~\ref{fig9}. We can see there that, upon
increasing $s$
from zero, the transition line separating the VL from the
BrG is nearly vertical, and that it is
only beyond $s\approx 5.0 K$ that the transition
line bends, while still retaining a quite appreciable $T$ dependence of the
value of $s$ at which the transition occurs. The transition line trends
towards being parallel to the $T$-axis at lower temperatures, indicating that the
BrG phase is thermodynamically 
stable only if the pinning strength is lower than a ``critical''
value which appears to be somewhat higher than $12.0K$.
Our result for the ``critical'' value of $s$ is in good agreement with an 
analytic estimate~\cite{vin} of this quantity for layered superconductors 
in the highly anisotropic limit where the electromagnetic interaction 
between vortices on different layers dominates over the interaction due to 
Josephson coupling.
The
general behavior of the transition temperature as a function of 
$s$ is in agreement with theoretical 
predictions~\cite{gautamprl,goldsc}. The shape of the phase boundary in 
Fig.~\ref{fig9} is also very similar to that found in other numerical
studies~\cite{nh,teitel2} of the BrG--VL transition.
The well-known result~\cite{review}
that the effective strength of pinning disorder in a sample with a fixed 
level of microscopic disorder increases 
with the  magnetic field $B$
suggests that the phase boundary in the $B-T$ plane would
have a shape similar to that in Fig.~\ref{fig9} with the $s$-axis replaced by
the $B$-axis. This would be consistent with  experimental
results~\cite{review}. 
However, caution should be exercised in translating our phase diagram in the
$s-T$ plane to one in the $B-T$ plane because 
the analogy between increasing $s$
at fixed $B$ and increasing $B$ for fixed disorder is not exact. 
As mentioned above, we find 
that the value of $D$, 
the average width of local density peaks at vortex positions 
as defined below Eq.~(\ref{width}), 
remains nearly constant along the phase boundary. This
supports  approximate analytic 
calculations~\cite{goldsc,nelson} based on the assumption that the 
melting transition of the BrG phase occurs at a constant value of the Lindemann
parameter ${\mathcal{L}}$. However, our results also show that if 
the quantity $D_{net}$, 
which measures the rms displacement of vortices from 
the average position of the
vortex line to which they belong (see discussion
in the paragraph following that including Eqn.~(\ref{width})), 
is used to define ${\mathcal{L}}$,
then it would not remain constant along the phase boundary -- it would
increase by about 15\% as the disorder strength is increased from zero to
$12.0 K$. Since the vortex positions on different layers are close to registry
in the BrG phase, while the interlayer correlations in the vortex positions
are very weak in the VL phase (see Figs.~\ref{fig1} and \ref{fig3}), the
BrG--VL transition may be thought of as a ``decoupling'' transition~\cite{glaz}
at which the correlation between vortex positions on different layers drops
abruptly.

We do not find evidence for  a VG phase 
thermodynamically distinct from the high-temperature VL 
in the range of pin strengths and concentrations studied. 
The VL minimum
obtained at relatively high temperatures evolves continuously as $T$
decreases. 
The degree of localization
of the vortices, measured by the typical heights of the local density peaks
at the free-energy minimum, does increase as $T$ is reduced, 
but the change is smooth.  
We found no evidence of
a phase transition between a weakly pinned and a strongly pinned distinct
VL states.
This leads us to classify {\it all}
disordered minima at low temperatures and relatively large pinning strengths
as {\it strongly pinned VL}, to distinguish them from the more liquid-like
disordered minima obtained for higher $T$ and smaller $s$. 
We have carried out various
thermal ``cycling'' procedures (such as warming up the disordered minimum 
obtained by quenching a liquid-like initial configuration to a low temperature,
and cooling the VL minimum obtained at a high temperature) for various values
of $s$ to look for local minima that are different from the BrG and VL ones. 
We did not find (see also below)
any other minimum that could be classified as a VG that is structurally 
and thermodynamically different from the VL.
Thus,
we conclude that our system does not exhibit a distinct equilibrium VG phase
in the region studied.
Since the interaction between straight vortex lines is screened in the model
considered here, such a conclusion 
would be  consistent with the expectation~\cite{bokil}
that a VG phase does nor exist in three dimensions if electromagnetic screening
of inter-vortex interactions is taken into account. However, 
such a drastically general inference would
be unwarranted. 
Also, our mean-field calculation that considers only the density variables
can not rule out the occurrence of a continuous VL-VG transition with subtle
characteristics (such as the continuous transition suggested by the
experimental results of Ref.~\onlinecite{zeldov}, 
for which structural signatures and the nature of the order 
parameter are particularly hard
to identify), 
or one in which the
phase of the superconducting order parameter (not included in our
treatment) plays a crucial role.

\begin{figure}
\includegraphics [scale=.66] {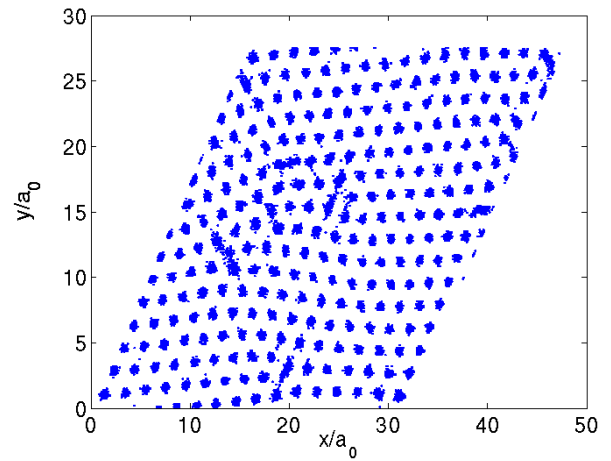}
\caption{\label{fig10} (Color online). Intermediate non-equilibrated multi-domain state
obtained at $s=9.0K$, $T=17.0 K$ as explained in the text. 
As in Fig.~\ref{fig1},
the average position of each vortex is represented by a dot.}
\end{figure}

We mentioned in Section \ref{intro} that earlier studies have suggested the
occurrence of other  phases for the system considered here. These 
include a ``vortex slush'' phase~\cite{nh}  predicted to exist at 
intermediate values of $T$ just above the boundary of the the BrG phase in 
the $s-T$ plane, and a
``domain glass'' phase~\cite{gautam}  expected  between the
BrG and VL phases for relatively 
small values of $s$. It has been shown~\cite{teitel}
that the vortex slush ``phase'' found in the simulation of Ref.~\onlinecite{nh}
corresponds to vortex configurations  nearly crystalline in each layer,
but with the vortices on different layers not
being in registry:  
the orientation of the reciprocal lattice vectors that describe the
crystalline order on different layers varies from layer to layer. The domain
glass phase\cite{gautam} is supposed to consist of 
large crystalline domains. We have looked for the 
existence of free-energy minima that may represent such
phases. In our attempts, we artificially made the VL minimum
unstable towards the formation of crystalline structures by 
multiplying the $C_{mn}(r)$ used as input in our calculations by a
factor of 1.05, which makes the structure factor $\tilde{S}({\bf k},k_z) 
\equiv 1/[1-\rho_0 C({\bf k},k_z)]$ {\it negative} 
for $k_z=0$ and $k$ close to the
magnitude of the smallest reciprocal lattice vectors of the 
triangular vortex lattice.
Our expectation was that the system would 
then converge to one of the other kinds of
minima mentioned above, if they existed. 
In these calculations, we started
with liquid-like initial conditions and quenched to within the 
region (low $T$, moderately large values of $s$) where these phases are 
expected. The minimization
routine was run for a large number of iterations with the modified $C_{mn}(r)$,
so that the instability of the VL minimum resulted, indeed, in the formation of
locally crystalline structures on the layers. Configurations obtained at
different stages of this run were then used as inputs in minimization runs 
in which we slowly eliminated the modification in $C_{mn}(r)$ so as to 
recover the original, physical interaction.
In all such runs,
the system converged to the BrG state but very slowly: the number 
of iterations it took for convergence was over 
twenty times larger than that in the
ordinary case. The
system lingered for a very large number of iterations in a multi-domain
state. The ultimate convergence was always
to minima with the vortices on different layers in registry,
even when the vortices were very much out of registry in the intermediate
configuration. An example is shown in Fig.~\ref{fig10}, at $s=9.0 K$, 
$T=17.0 K$ where
the vortex positions for an intermediate state with multi-domain
structure are shown in a plot
similar to those in Fig.~\ref{fig1}. The converged state, however, 
turned out
to be a single domain BrG very similar to that shown in the first two
panels of Fig.~\ref{fig1}. 

These results argue against the existence of the other phases suggested in
earlier studies, 
at least in the region of pinning strength and concentration 
studied, as thermodynamic equilibrium
phases. 
However, that the system remains in a polycrystalline 
state for a large number of iterations in the minimization process suggests
that such states may be stabilized through  changes in the model parameters
or may correspond to long-lived transient situations.
Further investigations of this possibility would be interesting. 
On the other hand, the reported
observation of this phase in simulations
might result from
using a small system size or insufficient equilibration, or both.
In our samples, the number of layers and the number of 
vortices on each layer are substantially larger than those in the simulations
of Ref.~\onlinecite{nh}. Also, the interlayer interaction in our model is
rather weak, since we consider only the electromagnetic interaction between 
vortices on different layers. The system nevertheless  converges
to a BrG minimum with vortices on different layers in  registry,
which implies that a vortex slush in which vortices form crystalline
arrangements on the layers, but the arrangements on different layers are not
in registry is rather unlikely to occur as a thermodynamic phase.  
 

\subsection{Distribution of local magnetic induction}

\begin{figure}
\includegraphics [scale=0.7] {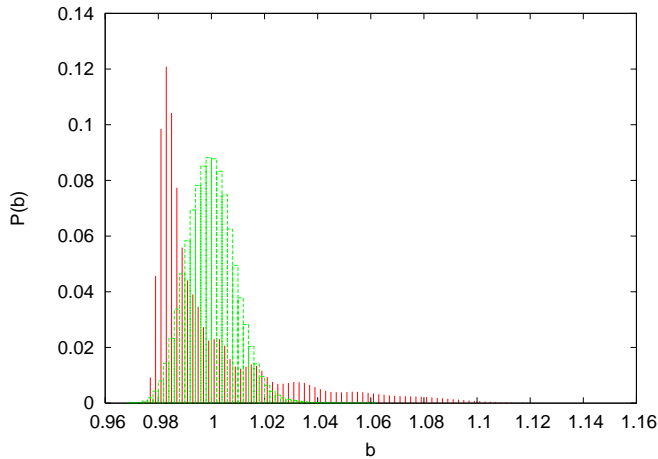}
\caption{\label{fig11} (Color online)
The probability distribution for fields $P(b)$ in terms
of the dimensionless field $b$ at $T=18.4K$ and $s=5.0 K$, which  
is very near the
melting point. The (red) solid lines are for the BrG state and the (green)
blocks are for the VL state.}
\end {figure}

Finally, we give  a brief discussion of our results for the distribution
of the local, microscopic, magnetic induction in the superconductor.
This is a quantity of considerable experimental interest, as it
can be measured directly in $\mu$SR experiments\cite{musrrev}. 
The
probability distribution of the local magnetic induction is easily 
computed\cite{musr} from the vortex density distribution
at the minima. We consider the the component of the magnetic induction
perpendicular
to the layers. The Fourier transform of this
$z$-component  due to a single pancake vortex at
the origin is given by
\cite{musr}
\begin{equation}
b_z({\bf k},k_z) = \frac{d \Phi_0}{1+\lambda^2 k^2 + \lambda^2 k_z^2},
\label{field1} 
\end{equation}
where $d$ is the spacing between the layers and $\lambda$ the penetration
depth in the layer plane. If the typical time scales for 
the dynamics of density fluctuations 
in the state  the vortex system is in
are much shorter than the characteristic 
time scale of muon spin precession, the muons see a broadened 
(time-averaged) density distribution of the vortices. 
The time-averaged local magnetic field $B_z$ is then obtained through 
a convolution of the local density at a free-energy minimum with the 
expression for $b_z$ given in Eq.(\ref{field1}). If the dynamic 
time scales for such fluctuations 
are much longer than
the precession time of the muon spins, then the convolution is to be 
done~\cite{musr} using a local density consisting of two-dimensional
$\delta$-functions located at the local density peaks at the 
free-energy minimum. The time scales are beyond the scope
of our time averaged calculation, but both methods give very similar results
in any case.

We have used these procedures to calculate the distribution of $B_z$ for
different free-energy minima. 
In our discretized system, the convolution is done using
discrete Fourier transforms, leading to the values of $B_z$ at the 
computational mesh points. These values are binned to yield histograms for
the distribution of $B_z$. The results are 
best represented in terms of the dimensionless field
$b\equiv B_z/B$. An example is given in Fig.~\ref{fig11} at a point on the
melting curve. 
The two states clearly have different probability
distributions $P(b)$, with that corresponding to the VL state being
narrower and much more symmetric. The distribution for the BrG state shows
a long ``tail'' in the large-$b$ side. 
All these features are very 
similar to experimental results~\cite{musrrev,lee1}.
The results shown in Fig.~\ref{fig11} were obtained using the locations
of the vortices at the minima -- the distributions obtained from 
the time-averaged local density have the same shape, but
are somewhat narrower in the VL phase. 

\begin{figure}
\includegraphics [scale=0.7] {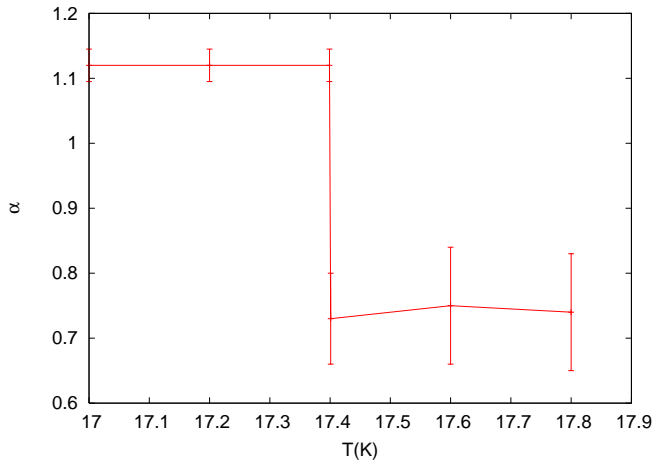}
\caption{\label{fig12} (Color online). The parameter $\alpha$ characterizing the degree of
asymmetry of the probability distribution $P(b)$ (see Figure \ref{fig11})
plotted as a function of $T$ at $s=10.0 K$. The result plotted at each 
temperature is the value
for the equilibrium state at that temperature, averaged over 
pin configurations.}
\end{figure}

The degree of asymmetry of $P(b)$ can be conveniently characterized\cite{musrrev} by
introducing a dimensionless
parameter $\alpha$ defined as the ratio of the cube root
of the third moment of $P(b)$ to the square root of its second moment. This
parameter changes discontinuously at melting. An example is shown in
Fig.~\ref{fig12} where $\alpha$ 
is plotted
as a function of $T$ at constant $s=10.0 K$. Results are  given for the
equilibrium state at each temperature and  
averaged over four pin configurations. It turns
out that for the VL state (but not for the BrG), the results
for $\alpha$ vary appreciably from one pin configuration
to another (the only quantity we have found that
does so) and also depending on which of the two methods
(local density peaks or full density) is used.  We have
therefore averaged also over the two methods.  
The resulting error bars are still larger in the VL
state as the distribution in that case is close to being symmetric
and rather small changes in the shape of $P(b)$ lead, in that
situation, to considerable changes in $\alpha$. The results for $\alpha$ 
are quantitatively similar to those of $\mu$SR experiments~\cite{lee1} on the
BrG-VL transition in BSCCO induced by increasing the magnetic field at a
fixed temperature. A similar discontinuity in the value of $\alpha$ is
also observed~\cite{lee2} at the melting of the BrG phase upon increasing the
temperature at fixed $B$. However, the value of $\alpha$ in the VL phase is
found~\cite{lee2} to be negative in that case. The reason for the negative 
value of $\alpha$ is not completely clear. It has been attributed to effects of 
sample geometry~\cite{musrrev} and it has been argued~\cite{divakar} 
that a negative $\alpha$
indicates the presence of a multi-domain state. Since effects of sample
geometry are not
present in our calculations with periodic boundary conditions and we do not
find any multi-domain state, it is perhaps not surprising that our calculations
do not yield negative values of $\alpha$. 

\section{Conclusions}
\label{summary}

We have studied in this paper the phase diagram of the superconducting vortex
lattice in a layered superconductor, in the presence of both a magnetic field
normal to the layers and of random atomic-scale point pinning centers. We
have considered the strongly anisotropic case where the interactions are 
of two-body form and the free energy can then be written  as a functional of
the time averaged vortex density. Numerical minimization of the discretized
free energy then yields the pancake vortex density distribution at the minima.

The values of the free energy at the minima can be plotted to locate the
phase transitions. Our main result is the phase diagram (Fig.~\ref{fig9})
in the temperature -- pin strength plane, where we find a line of first-order
phase transitions between a Bragg glass and a disordered phase.
The density  distributions at the minima can then be analyzed to obtain
any desired correlation functions. They can also be used to directly visualize
the vortex lattice in real space and to
obtain the distribution of the microscopic local magnetic
induction in the 
superconductor. From an extensive analysis of such
results we conclude that the disordered phase is to be identified as a pinned
vortex liquid. While one can distinguish between strong  and  
weak pinning regimes,
the crossover between them is completely smooth.

As discussed in detail in the previous section, our results are 
consistent with those of experiments and simulations, the only possible
exception being that we do not find, 
in the region of pinning strength and concentration studied, 
 any sign of additional 
``vortex glass'',
``vortex slush''
or ``domain glass'' phases the existence of which has been made at least
plausible by results obtained e.g. in Refs.~\onlinecite{zeldov,
nh,gautam} respectively. We have 
ourselves found~\cite{prbd,us1,us2} 
a polycrystalline Bose glass phase in
the case of columnar pinning through the use of the same methods 
as in this work. 
Such phases may exist under other conditions, specifically
for a much smaller concentration of considerably stronger  defects
which would favor polycrystallinity much more than the situation studied
here, where 
we have chosen a value of the concentration
$c$ of atomic defects which, although small in the atomic scale, is in a sense
large with respect to the vortex lattice: there are many pinning sites per
vortex lattice unit cell. The results for columnar pinning, where the 
concentration is much smaller but the pinning strength larger, make it possible
to conjecture that there is a regime of larger $s$ but smaller $c$ where
additional phases may  exist. Further work in that direction is clearly
worthwhile.

In a paper~\cite{rodri} that appeared after
this one was originally submitted,
it has been suggested, from an analysis of a disordered $xy$ model in
which electromagnetic interactions between vortices is neglected, 
that the BrG phase should 
become unstable (no matter how weak the pinning) to disorder
in the direction normal to the layers at a value of $B$ smaller than that
considered here. Our results for the correlation
function $g(0,n)$ as a function of interlayer separation $n$ in the ordered
phase (discussed in the text in connection  with Fig.~\ref{fig3}), indicate
no decay in this quantity as a function of $n$ in the region 
$3 < n \lesssim N_L/2$.
Further, as mentioned above, our phase diagram data quantitatively agree 
with analytic calculations~\cite{vin} in which the softening of the tilt 
modulus at large wavelengths was included. These results strongly argue 
for the actual stability of the BrG phase, in agreement with  
previous numerical studies~\cite{nh,teitel2} of the disordered $xy$ 
model. However,
even though our samples are already, as noted, much thicker 
than those considered in earlier numerical studies
\cite{nh,teitel2}, the possibility that the BrG phase would 
disorder in the transverse
direction over a length scale much
larger than $\lambda$ can not be mathematically ruled
out by our numerical results.

%


\begin{acknowledgments}
This work was supported  in part by NSF (OISE-0352598) and by
DST (India).
\end{acknowledgments}


\end{document}